\documentclass[a4paper,11pt]{article}
\pdfoutput=1 

\usepackage{jcappub} 

\usepackage[T1]{fontenc} 

\usepackage{latexsym}    
\usepackage{graphicx}   
\usepackage{verbatim}   
\usepackage[normalem]{ulem} 
\usepackage[dvipsnames]{xcolor}     
\usepackage{tikz}       
\usetikzlibrary{shapes}
\usepackage{subfigure}  
\usepackage{multirow}
\usepackage{hhline}
\usepackage{url}
\usepackage{hyperref}   
\usepackage{tensor}     
\usepackage{amssymb}
\usepackage{enumerate}	
\usepackage{braket}		
\usepackage{physics}
\newcommand{\REV}[1]{#1}
\newcommand{\COMREV}[1]{}
\newcommand{\HWC}[1]{#1}
\newcommand{\COMHWC}[1]{}

\newcommand{\chicrit}{\sqrt{2/(3\kappa^2)}}

\title{\boldmath Observational constraints on 3-forms  dark energy}

\author[a,b,c]{Mariam Bouhmadi-López,}
\author[d]{Hsu-Wen Chiang,}
\author[b,c]{Carlos G. Boiza,} 
\author[e,f]{Pisin Chen}

\affiliation[a]{IKERBASQUE, Basque Foundation for Science, 48011, Bilbao, Spain}
\affiliation[b]{Department of Physics, University of the Basque Country UPV/EHU, P.O. Box 644, 48080 Bilbao, Spain}
\affiliation[c]{EHU Quantum Center, University of the Basque Country UPV/EHU, P.O. Box 644, 48080 Bilbao, Spain}
\affiliation[d]{Department of Physics, Southern University of Science and Technology, Shenzhen 518055, China}
\affiliation[e]{Leung Center for Cosmology and Particle Astrophysics,
National Taiwan University, Taipei 10617, Taiwan}
\affiliation[f]{Department of Physics and Graduate Institute of Astrophysics,
National Taiwan University, Taipei 10617, Taiwan}

\emailAdd{mariam.bouhmadi[at]ehu.eus}
\emailAdd{jiangxw[at]sustech.edu.cn, b98202036[at]ntu.edu.tw}
\emailAdd{carlos.garciab[at]ehu.eus}
\emailAdd{pisinchen[at]phys.ntu.edu.tw}

\abstract{3-forms are natural candidates for describing the late-time accelerated expansion of the Universe, as they can inherently reproduce a positive cosmological constant when lacking an evolving potential. When such a potential is present, a 3-form field may exhibit either quintessence-like or phantom-like behaviour. In this paper, we consider a late-time effective dark energy model described by a 3-form with a Gaussian potential, stable during the dark-energy-dominated era. We constrain this model observationally by performing a Markov Chain Monte Carlo (MCMC) analysis employing a comprehensive cosmological dataset, including Planck PR4 cosmic microwave background (CMB) data, DESI DR1 baryon acoustic oscillation (BAO) measurements, Pantheon+ Type Ia supernovae data, low-$z$ Cepheid calibrators, and DES Y1 large-scale structure observations. We demonstrate that the 3-form model successfully increases the predicted Hubble parameter of CMB and BAO data from $67.89\pm0.36{\rm km/s/Mpc}$ of $\Lambda$CDM model to $68.29^{+0.56}_{-0.61}{\rm km/s/Mpc}$ \COMREV{without fine-tuning of the model parameters}\REV{by approaching the potential peak at the right time}, thus \REV{mildly }reducing the tension with the late-time observation. \COMREV{Furthermore, we verify the sub-dominance of the 3-form field perturbation via both analytical and numerical analyses. Thus}\REV{Overall}, the 3-form field \COMREV{does }serve\REV{s} as a promising candidate of phantom-like dark energy from both theoretical and observational points of view.
}

\begin{document}
\maketitle
\flushbottom

\section{Introduction}
\label{sec:intro}


Advances in observational cosmology are providing an unprecedented wealth of data, enabling increasingly precise description of the cosmic evolution as a probe for fundamental theories of gravitation. A key unresolved issue is identifying the mechanism responsible for the accelerated expansion of the Universe observed in the current epoch. This phenomenon was first indicated nearly thirty years ago through studies of Type Ia supernovae \cite{SupernovaSearchTeam:1998fmf,SupernovaCosmologyProject:1998vns,Bahcall:1999xn} and has since been independently confirmed by complementary probes, including measurements of the cosmic microwave background (CMB) \cite{Bennett2003,Komatsu2011,Planck2013,Planck2018}, baryon acoustic oscillations (BAO) \cite{Eisenstein2005,Cole2005,Percival2010}, and analyses of the large-scale distribution of matter \cite{Tegmark2004,Reid2010}.

The late-time acceleration of the Universe is generally ascribed to dark energy; i.e. an unknown component with negative pressure that dominates the cosmic energy budget. The simplest explanation invokes a cosmological constant, giving rise to the concordance $\Lambda$CDM model; however, this approach entails significant theoretical challenges, including the fine-tuning and coincidence problems. These challenges have motivated a variety of extensions in which dark energy evolves dynamically rather than remaining constant, including scalar field models such as quintessence \cite{Ratra:1987rm}, k-essence \cite{Armendariz-Picon:2000ulo}, 
axion-like DE models~\cite{Kamionkowski:2014zda,Emami:2016mrt,Chiang:2025qxg,Bouhmadi-Lopez:2025wxo}, 
kinetic gravity braiding \cite{Deffayet:2010qz,Pujolas:2011he,BorislavovVasilev:2022gpp,BorislavovVasilev:2024loq}, and also Horndeski models \cite{Kobayashi:2019hrl}. Other approaches include sign-switching cosmological constant models, in which the effective cosmological term changes sign over cosmic time \cite{Akarsu:2021fol, Akarsu:2022typ, Akarsu:2023mfb, Bouhmadi-Lopez:2025ggl, Bouhmadi-Lopez:2025spo}, and perfect-fluid descriptions of dark energy with parametrised equations of state \cite{Kamenshchik:2001cp,Bouhmadi-Lopez:2004mpi,Bouhmadi-Lopez:2006fwq}. Alternatively, the observed cosmic acceleration can be attributed to modifications of gravity rather than to a new dark energy component. In this approach, the late-time dynamics of the Universe arise from extensions or generalisations of General Relativity, offering an alternative explanation for the accelerated expansion while potentially alleviating some of the theoretical challenges associated with $\Lambda$CDM \cite{CANTATA:2021asi}. Some examples are $f(\mathcal{R})$ gravity \cite{Capozziello:2011et,Nojiri:2010wj,Nojiri:2017ncd,Morais:2015ooa},
$f(\mathcal{T})$ models \cite{Bengochea:2008gz,Ferraro:2006jd,Cai:2015emx},  and  $f(\mathcal{Q})$ theories~\cite{BeltranJimenez:2018vdo,BeltranJimenez:2019tme,Ayuso:2020dcu,Boiza:2025xpn,Ayuso:2025vkc},  among others. Collectively, these models provide a versatile framework for exploring physics beyond $\Lambda$CDM and for confronting the late-time acceleration of the Universe with observational data.

Beyond the theoretical challenges outlined above, the $\Lambda$CDM model exhibits persistent observational tensions revealed by high-precision cosmological data. Foremost is the discrepancy between the Hubble constant, $H_0$, inferred from cosmic microwave background (CMB) observations by the \textit{Planck} satellite \cite{Rosenberg:2022sdy} and that obtained from local distance-ladder measurements, such as Cepheid-calibrated Type~Ia supernovae, most notably by the SH0ES collaboration \cite{Riess:2020fzl, Brout:2022vxf}. This $H_0$ tension, remaining at a significance exceeding $5\sigma$, may hint at physics beyond the standard cosmological model. A far milder discrepancy concerns the amplitude of matter fluctuations, expressed by the parameter $S_8$, for which large-scale structure probes—including weak-lensing surveys such as KiDS \cite{Wright:2025xka}, DES \cite{DES:2021wwk,DESCollaboration:2025udj}, and HSC \cite{Dalal:2023olq}—yield systematically lower values than those inferred from \textit{Planck} within the $\Lambda$CDM framework \cite{Rosenberg:2022sdy}. Though this tension nowadays is almost gone. Collectively, these tensions may signal limitations in our understanding of cosmic evolution or in the underlying assumptions of the standard model. Moreover, the latest results from DESI have rekindled the debate on the nature of dark energy, raising the possibility that it may not be a simple cosmological constant but rather a dynamical component that evolves over cosmic time. This interpretation arises from subtle deviations observed in the expansion history and structure-growth measurements, which, although still consistent with the $\Lambda$CDM model within uncertainties, suggest mild preferences for a time-varying dark energy equation of state (EoS)\cite{DESI:2025fii}.

In this paper, we focus on 3-form fields, which have emerged as compelling candidates for modelling dark energy \cite{Koivisto:2009fb,Koivisto_2013,Morais:2016bev, Bouhmadi-Lopez:2016dzw,Morais:2017vlf} and which naturally arise in higher-dimensional frameworks such as string theory and supergravity \cite{Aurilia:1980xj}. By inherently generating an effective negative pressure, these fields can account for the accelerated expansion of the Universe in its late-time evolution \cite{Koivisto:2009ew}. In particular, under specific conditions; such as a constant or vanishing potential, the dynamics of a 3-form field can closely replicate those of a cosmological constant \cite{Duff:1980qv}, thereby providing a natural mechanism to explain the inflationary era and the observed dark energy component \cite{Koivisto:2009fb,Morais:2016bev}.

In fact, 3-form fields have proven to be robust dynamical components in cosmological model building, with applications spanning from the inflationary epoch to the late-time acceleration of the Universe. Within the inflationary framework, scenarios involving a single 3-form field minimally coupled to gravity were first formulated and systematically analysed in \cite{Koivisto:2009fb,Koivisto:2009ew} (cf. also \cite{Germani:2009iq}). The issues of ghost and Laplacian instabilities were subsequently investigated in \cite{DeFelice:2012jt}, where the corresponding stability conditions were explicitly derived, while a consistent reheating mechanism was proposed in \cite{DeFelice:2012wy}. The phenomenological viability of these scenarios as early-Universe inflationary models has been examined in both single-field \cite{Mulryne_2012,Urban:2012ib} and multi-field \cite{Kumar:2014oka,SravanKumar:2016biw} realisations. Furthermore, brane-world inflationary models driven by a 3-form field confined to the brane hypersurface have also been proposed and analysed in \cite{Barros:2015evi} (see also \cite{Barros:2023nzr, Gordin:2023nsv}).

At late times, 3-form fields have been extensively investigated as candidates for dark energy, including scenarios involving interactions between dark matter and dark energy, where de Sitter attractors, quintessence, and phantom-like behaviour may naturally emerge depending on the specific form of the 3-form potential \cite{Morais:2016bev,Koivisto:2009fb,Koivisto_2013,Morais:2017vlf}. Moreover, such phantom-like behaviour has, in turn, facilitated the construction of wormhole solutions supported by 3-form fields \cite{Bouhmadi-Lopez:2021zwt} (cf. also \cite{Barros_2018}), as well as non-singular black hole configurations \cite{Bouhmadi-Lopez:2020wve} (see also \cite{Barros:2020ghz}). In addition, k-essence models have been extended to include 3-form dynamics \cite{daFonseca:2024boz}, and anisotropic geometries have recently been explored within this framework \cite{DeFelice:2025khe}.
Screening mechanisms for 3-form fields have been investigated, demonstrating that 3-form dark energy models can satisfy current Solar System constraints \cite{Barreiro:2016aln}. Static, spherically symmetric 3-form stars were analysed in \cite{Barros:2021jbt}. The quantum cosmology of 3-form fields has also been studied in \cite{Bouhmadi-Lopez:2018lly}. To the best of our knowledge, however, no observational constraints on dark energy models driven by 3-form fields have yet been established, a gap that the present work aims to address. Furthermore, we seek to investigate the phantom-like behaviour induced by 3-form fields as a potential means of alleviating current cosmological tensions.

In this work, we study a 3-form field model with a Gaussian potential, which gives rise to a dark energy component exhibiting phantom-like behaviour at late-time. This behaviour stems from the fact that the potential decreases with the square of the three-form field while remaining strictly positive \cite{Morais:2017vlf,Morais:2016bev} (cf. also Eq.(\ref{3form_w})). In addition, we place observational constraints on this model. Specifically, we perform comprehensive Markov Chain Monte Carlo (MCMC) analysis across various combinations of dataset drawn from a list of recently released measurements, including early-time, late-time, background and perturbative observations: \textit{Planck} 2018 low-$\ell$ CMB TTEE \cite{Planck:2019nip}, NPIPE PR4 \textit{Planck} CamSpec high-$\ell$ CMB TTTEEE \cite{Rosenberg:2022sdy}, NPIPE PR4 \textit{Planck} CMB lensing \cite{Carron:2022eyg,Carron:2022eum}, DESI DR1 BAO without fullshape analysis \cite{DESI:2024mwx,DESI:2024lzq, DESI:2024uvr}, Pantheon+ SNe Type Ia without anchoring of the standardised absolute magnitude \cite{Brout:2022vxf}, local $H_0$ measurements \cite{Riess:2021jrx} and DES Y1 weak lensing dataset \cite{DES:2017myr}. In particular, the inclusion of CMB spectrum data allows us to assess whether the \COMREV{3-form field perturbation remains well-controlled during}\REV{matter perturbation evolves differently under the modified background} cosmic evolution.

These MCMC analyses then allow us to statistically compare the 3-form model endowed with a Gaussian potential to the $\Lambda$CDM setup, via either direct comparison of the relevant parameters such as $H_0$, $\Omega_{m0}$ and $S_8$, or via statistical probes recommended in prior researches \cite{DES:2020hen, Raveri:2019gdp, PhysRevD.100.023512}. Parameters not directly related to the cosmological tension are checked as well for better understanding of the 3-form model. Finally, we utilise CAMB Boltzmann solver \cite{Lewis:1999bs,Howlett:2012mh} to generate CMB power spectrum, $f\sigma_8$ and matter power spectrum. The results are compared with observations \cite{eBOSS:2020lta, Turnbull:2011ty, Achitouv:2016mbn, Beutler:2012px, Feix:2015dla, BOSS:2016wmc, BOSS:2013eso, Blake:2012pj, Nadathur:2019mct, BOSS:2013mwe, Wilson:2016ggz, eBOSS:2018yfg, Okumura:2015lvp, Reid:2009xm, eBOSS:2018qyj} to determine whether there are any potential effect on the large scale structure formation.

Our paper is organised as follows. In Section \ref{sec:model}, we review the cosmology of 3-form fields, starting from the gravitational action of a minimally coupled 3-form with an arbitrary potential. In Section \ref{sec:gaussian}, we motivate the use of a Gaussian potential as a means of describing the late-time acceleration of the Universe, considering a Universe filled with matter and radiation, in addition to 3-form dark energy.
We then perform a thorough and systematic dynamical analysis that extends the previous work conducted by one of the authors  \cite{Morais:2016bev,Morais:2017vlf}.
In Section \ref{sec:result}, we present our observational analysis, including the prior selection, spectrum analysis, and the statistical comparison between the 3-form model and $\Lambda$CDM model. We also provide theoretical analysis of the 3-form dynamics. We conclude in Section \ref{sec:conclusion} on the pros and cons of the 3-form field with a Gaussian potential as a dark energy model. Finally, in Appendix \ref{app:CM-ours} we summarise the Centre Manifold approach used in Section \ref{sec:gaussian}, while in Appendix \ref{sec:stat} we define the cosmological probes employed in Section \ref{sec:result}.

\section{3-forms gravity}
\label{sec:model}

\subsection{3-form Fields in General Relativity}

We begin by defining a general action for a 3-form field $A_{\mu\nu\rho}$; i.e. a rank-three antisymmetric
tensor field,  minimally coupled to gravity and with a given potential $V$ which can be written as 
\footnote{Throughout this paper, Greek indices will be used for four-dimensional quantities, while Latin indices will denote three-dimensional spatial quantities.} 
\begin{align}
	\label{N3-form action}
	S
	 = \int \mathrm{d}^4x \, \sqrt{-g}\,\, \left(\frac{1}{2\kappa^2}\, R + \mathcal{L}\right),
\end{align}
where the Lagrangian density $\mathcal{L}$ reads
\begin{align}\label{L3-form Lagrangian}
	\mathcal{L}
	=
		-\frac{1}{48} F^{\mu\nu\rho\sigma} F_{\mu\nu\rho\sigma}
		-V\left(A^{\mu\nu\rho}A_{\mu\nu\rho}\right)
	\,.
\end{align}
In the action (\ref{N3-form action}), $g$ is the determinant of the metric, $\kappa^2=8\pi G$ where $G$ is the gravitational constant, and in the Lagrangian density (\ref{L3-form Lagrangian}) $F_{\mu\nu\rho\sigma}$ is the strength tensor of the 3-form, defined as \cite{Germani:2009iq,Koivisto:2009ew,BeltranAlmeida:2018nin}
\begin{align}
	\label{N3f-Maxw-1}
	F_{\mu\nu\rho\sigma}\equiv4\nabla_{[\mu}A_{\nu\rho\sigma]}
	=
	\nabla_{\mu}A_{\nu\rho\sigma}
	-\nabla_{\sigma}A_{\mu\nu\rho}
	+\nabla_{\rho}A_{\sigma\mu\nu}
	-\nabla_{\nu}A_{\rho\sigma\mu}
	\,,
\end{align}
where the square brackets is the usual notation for anti-symmetrisation.

Minimising the action \eqref{N3-form action} with respect to variations of the 3-form field yields the equations of motion (EoM) for the 3-form (cf. ref. \cite{Germani:2009iq,Koivisto:2009ew})
\begin{align}
	\label{general_eq_motion}
	\nabla_{\sigma}{F^{\sigma}}_{\mu\nu\rho}
		-12\frac{\partial \,V}{\partial 	\left(A^2\right)} A_{\mu\nu\rho}=0
		\,,
\end{align}
where $A^2$ stands for $A^{\alpha\beta\gamma}A_{\alpha\beta\gamma}$.

Similarly, minimising the action \eqref{N3-form action} with respect to variations of the metric lead to 

\begin{align}
	\label{Einstein_eq_motion}
	G_{\mu\nu} = \kappa^2 \,\, T_{\mu\nu}.
\end{align}
In the former equation, $G_{\mu\nu}$ stands for Einstein tensor while $T_{\mu\nu}$ is  the energy-momentum tensor of the 3-form. 
The latter can be derived from the action \eqref{N3-form action} and is given by \cite{Koivisto:2009ew,Koivisto:2009fb}
\begin{align}
	\label{energy_momentum_3form}
	{T}_{\mu\nu}
	\equiv
	\frac{-2}{\sqrt{-g}}\frac{\partial \sqrt{-g}\mathcal{L}}{\partial g^{\mu\nu}}
	=&~
	\frac{1}{6}  {F}_{\mu}^{\phantom{\mu}\alpha\beta\gamma} F_{\nu\alpha\beta\gamma} 
	+6\frac{\partial \,V}{\partial \left(A^2\right)} {A}_{\mu}^{\phantom{\mu}\alpha\beta} A_{\nu\alpha\beta}
	-\left[
		\frac{1}{48}F^2
		+V \left(A^2\right)
	\right] g_{\mu\nu}
	\,.
\end{align}
 It can be shown that, for a vanishing or constant potential, the 3-form behaves as a cosmological constant \cite{Aurilia:1980xj,Duff:1980qv}. We will return to this point in more detail when discussing the cosmological framework. Moreover, given that the 3-form is minimally coupled to gravity and no interaction with dark matter is assumed, its energy–momentum tensor is conserved. In the remainder of this work, we assume that the three-form potential is always positive.
 
\subsection{Late-time Cosmology with 3-form Fields}

To describe the universe’s dynamics, we adopt a spatially flat Friedmann-Lema\^{i}tre-Robertson-Walker (FLRW) metric:
\begin{equation}\label{rwmetric}
   ds^2=-dt^2+a^2(t)d\bar{x}^2.
\end{equation}
Here, $t$ is the cosmic time and $a(t)$ represents the scale factor. We consider such a FLRW universe filled with matter, radiation, and a 3-form field, the latter acting as dark energy. Given the symmetry incorported by the FLRW space-time, the 3-form field depends only on the cosmic time, therefore only the space-like components will be dynamical%
\footnote{We can set all the non-dynamical components $A_{0ij}$ to zero \cite{Koivisto:2009ew}.
}%
, with its non-zero components given by \cite{Koivisto:2009ew,Koivisto:2009fb}
\begin{align}
	\label{NNZ-comp-1-1}
	A_{ijk}=a^{3}(t)\,\chi(t)\,\epsilon_{ijk}
	\,,
\end{align}
while the non-zero components of the strength tensor are (cf. Eq.(\ref{N3f-Maxw-1}))
\begin{align}
	\label{F_0ijk_comp}
	F_{0ijk} = a^{3}(t)\left[\dot\chi(t)+3H(t)\chi(t)\right]\epsilon_{ijk}
	\,.
\end{align}
In the former equations, $\epsilon_{ijk}$ is the standard 3-dimensional Levi-Civita symbol and $\chi(t)$ represents the comoving scalar degree of freedom corresponding to the 3-form field. Here, $H$ denotes the Hubble parameter, and a dot henceforth indicates differentiation with respect to cosmic time $t$. From Eq.~\eqref{NNZ-comp-1-1}, we obtain $A^{\mu\nu\rho}A_{\mu\nu\rho}=6\chi^{2}$, which allows the potential to be expressed as $V(\chi^2)$. Moreover, Eq.~\eqref{F_0ijk_comp} implies that $F^{\mu\nu\rho\sigma}F_{\mu\nu\rho\sigma}=-24\left(\dot{\chi}+3H\chi\right)^2$.
Therefore, the energy
density and pressure of the 3-form can be written as follows
\begin{align}
	\label{energy_3form}
	\rho_\chi =&~ -{T^0}_0 = \frac{1}{2}\left(\dot\chi+3H\chi\right)^2 + V
	\,,
	\\
	\label{pressure_3form}
	P_\chi =&~ \frac{1}{3}{T^i}_i = -\frac{1}{2}\left(\dot\chi+3H\chi\right)^2 - V + \chi V_{,\chi}
	\,.
\end{align}
Consequently, the Friedmann and Raychaudhi equations read
\begin{align}
	\label{NFriedm-1-1}
	H^{2}=\frac{\kappa^{2}}{3}\left[\rho_{\mathrm{m}} +\rho_{\mathrm{r}}+
		\frac{1}{2}\left(\dot{\chi}+3H\chi\right)^{2}
		+V
	\right]
	\,,
\end{align}
\begin{align}
	\label{NHdot-1-1}
	\dot{H}=-\frac{\kappa^{2}}{2}\left[\rho_{\mathrm{m}} +\frac43\, \rho_{\mathrm{r}}+\chi V_{,\chi}\right]
	\,,
\end{align}
respectively. 

Before proceeding, we briefly examine the equation of state (EoS) of the 3-form, which can be straightforwardly defined from Eqs.~(\ref{NFriedm-1-1}) and (\ref{NHdot-1-1}) and reads
\begin{align}
	\label{3form_w}
	w_\chi = \frac{P_\chi}{\rho_\chi} = -1 + \frac{\chi V_{,\chi}}{\frac{1}{2}\left(\dot\chi+3H\chi\right)^2 + V}
	\,.
\end{align}
As is evident from the preceding equation, when the 3-form potential is either constant or vanishing, its EoS takes the value $-1$. Given that the energy–momentum tensor of the 3-form remains conserved within the class of models considered, it follows that, in such scenarios, the three-form effectively behaves as a cosmological constant. Moreover, the EoS of the 3-form will always be smaller than $-1$ if the potential is a decreasing function of the square of the 3-form, thereby exhibiting a phantom-like behaviour. Conversely, the EoS will be greater than $-1$ if the potential is an increasing function of the square of the 3-form, in which case the field behaves analogously to a quintessence model.

Finally, let us move to EoM of the 3-form. Substituting Eqs.\eqref{NNZ-comp-1-1} and \eqref{F_0ijk_comp} in  Eq.\eqref{general_eq_motion}, yields the equation of motion for $\chi(t)$ 
\begin{align}
	\label{NDiff-syst-1-1}
	\ddot{\chi}+3H\dot{\chi}+3\dot{H}\chi+V_{,\chi}=0
	\,.
\end{align}
Using the Raychaudhuri equation \eqref{NHdot-1-1} to eliminate $\dot H$ in Eq.~\eqref{NDiff-syst-1-1} we find
\begin{align}
	\label{chi_eq_motion}
	\ddot{\chi}+3H\dot{\chi}+\left(1 - \left(\frac{\chi}{\chi_c}\right)^2\right)V_{,\chi}=\frac{3\kappa^2}{2}\left(\rho_{\mathrm{m}} +\frac43\, \rho_{\mathrm{r}}\right)\chi
	\,.
\end{align}
where $\chi_\mathrm{c}\equiv\chicrit$. The preceding equation is particularly illustrative, as it shows that when matter and radiation redshift away, the system admits two types of critical points: one associated with the extrema of the potential, and another corresponding to $\chi=\chicrit$. Naturally, for a given potential, a thorough dynamical system analysis must be performed in order to identify all the critical points of the system and assess their stability. This is a task we address on the next section (cf. also, for example, \cite{Morais:2016bev, Bouhmadi-Lopez:2016dzw}).  

From this point onward, we focus on a three-form endowed with a Gaussian potential, which, being a decreasing function of the square of the field, exhibits phantom-like behaviour. We further assess its potential to alleviate the Hubble tension. As discussed in the next section, this class of potential also ensures the stability of three-form perturbations\REV{~at the dark-energy-dominated era}, leading to a well-behaved \REV{effective }dark energy model.

\section{A dark energy 3-form with a Gaussian potential}
\label{sec:gaussian}

\subsection{The model}

To obtain a phantom-like EoS with a 3-form while avoiding ghosts, the potential must decrease with $\chi^2$ so that $\chi V_{,\chi}<0$ and $w_{\COMREV{X}\REV{\chi}}<-1$ (see \eqref{3form_w}), while \COMREV{also yielding a healthy sound speed}\REV{a speed of sound \COMHWC{$c_s$ }in between 0 and 1 is also required to avoid the gradient instability}. We remind \COMREV{at}\REV{in} this regard\COMREV{s} that the speed of sound for a 3-form minimally coupled to gravity reads \cite{Koivisto:2009fb}

\begin{equation}
\label{cs2eq}
c_s^2=\frac{\chi V_{,\chi\chi}}{V_{,\chi}}.
\end{equation}
\COMREV{These requirements are}\REV{The ghost stability is} naturally satisfied by the Gaussian form \cite{Koivisto:2009fb,Morais:2016bev}

\begin{equation}
V(\chi)=V_\ast\,\exp\!\left(-\frac{\xi\kappa^2}{6}\chi^2\right), \qquad V_\ast,\xi>0,
\end{equation}
which has a single maximum at $\chi=0$ and decreases monotonically for large $|\chi|$. In this case,

\begin{equation}
c_s^2=1-\frac{\xi\kappa^2}{3}\chi^2.
\end{equation}
Thus the field exhibits phantom behaviour whenever $\chi\neq0$ (c.f. \eqref{3form_w}), while $c_s^2$ remains positive and subluminal near the origin (see \eqref{cs2eq})\REV{, but not beyond the inflection point $V_{,\chi\chi} = 0$ at 1-$\sigma$ away from the centre of the Gaussian\footnote{\REV{\HWC{The Gaussian potential can be expressed as $\exp \left[ - \left( \chi / \sigma \right)^2 / 2 \right]$ where $\sigma = \sqrt{3\xi^{-1}} \kappa^{-1}$, and borrowing the notion from statistics we describe the inflection point to be at 1-$\sigma$, i.e., $\chi = \pm\sqrt{3\xi^{-1}} \kappa^{-1}$.
}}}}
. Since the dynamics of a 3-form asymptotically confine\REV{s} the field to the interval $[-\chi_c,\chi_c]$, viability requires that this interval lies entirely within the region where $c_s^2>0$, namely \cite{Morais:2016bev}

\begin{equation}
\chi\in\left[-\sqrt{\frac{9}{2\xi}}\,\chi_c,\;\sqrt{\frac{9}{2\xi}}\,\chi_c\right].
\end{equation}
This leads to the condition

\begin{equation}
0<\xi<\frac{9}{2}\,,
\end{equation}
which ensures that no perturbative instabilities appear once the system settles inside $[-\chi_c,\chi_c]$\REV{, thus putting an upper bound on the growth factor of the 3-form perturbation due to the gradient instability. However, we emphasize that the perturbation is inevitably unstable in the early universe, and therefore one should treat this 3-form theory as an effective description valid only at late time. We will estimate the growth factor, and thus at which epoch the 3-form field description becomes reliable at the perturbative level, once we have a grasp on the behaviour of the background 3-form field}.

To analyse the stability of the Gaussian 3-form model, we adopt a dynamical system approach. In addition to the 3-form field, we include a radiation component with EoS $w_\mathrm{r}=1/3$ and a pressureless matter component with $w_\mathrm{m}=0$. Both satisfy the usual conservation laws,

\begin{equation}
\dot\rho_\mathrm{r}+4H\rho_\mathrm{r}=0, \qquad \dot\rho_\mathrm{m}+3H\rho_\mathrm{m}=0.
\end{equation}
The total EoS of the system is defined in the usual way as
\begin{equation}
w_{\mathrm{tot}} = \frac{p_{\chi} + p_\mathrm{r} + p_\mathrm{m}}{\rho_{\chi} + \rho_\mathrm{r} + \rho_\mathrm{m}},
\end{equation}
where $(\rho_\mathrm{i},p_\mathrm{i})$ denote the energy density and pressure of each component.

\subsection{Dynamical system analysis}
\label{sec: dyn_var}

Following \cite{Bouhmadi-Lopez:2016dzw}, we employ compact dimensionless variables that map the dynamics into a bounded phase space. In particular, we define \footnote{We note that the variables $y$ and $s$ were originally proposed in \cite{Koivisto:2009fb}.}

\begin{equation}
\frac{v}{1-v^2}=\frac{\chi}{\chi_c}, 
\qquad 
y=\frac{\dot\chi+3H\chi}{3H\chi_c},
\qquad 
\mathfrak{h}=\frac{(H/H_0)^2}{1+(H/H_0)^2},
\qquad
r=\frac{\kappa\sqrt{\rho_\mathrm{r}}}{\sqrt{3}H},
\qquad
s=\frac{\kappa\sqrt{\rho_\mathrm{m}}}{\sqrt{3}H},
\end{equation}
with $-1\leq v \leq 1$, $-1\leq y \leq 1$, $0\leq \mathfrak{h} \leq 1$, $0\leq r \leq 1$, and $0\leq s \leq 1$. These variables automatically satisfy the Friedmann constraint

\begin{equation}
y^2+r^2+s^2+\frac{1-\mathfrak{h}}{\mathfrak{h}}\bar{V}\exp\left[-\frac{\xi}{9}\left(\frac{v}{1-v^2}\right)^2\right]=1,
\label{fconstraint}
\end{equation}
where $\bar{V}$ is a dimensionless variable defined as

\begin{equation}
\bar{V}=\frac{\kappa^2V_\ast}{3H_0^2}.
\end{equation}
In terms of the compact variables $(v,y,\mathfrak{h},r)$, the total EoS can be written explicitly as
\begin{equation}
\begin{aligned}
w_{\rm tot}(v,y,\mathfrak{h},r) =\;& 
- \left[ y^2 
+ \frac{1-\mathfrak{h}}{\mathfrak{h}}\,\bar V 
   \exp\!\left[-\frac{\xi}{9}\,\frac{v^2}{(1-v^2)^2}\right] \right] \\[6pt]
&- \frac{2\xi}{9}\,\frac{v^2}{(1-v^2)^2}\,
   \frac{1-\mathfrak{h}}{\mathfrak{h}}\,\bar V 
   \exp\!\left[-\frac{\xi}{9}\,\frac{v^2}{(1-v^2)^2}\right] 
+ \frac{1}{3}r^2 .
\end{aligned}
\label{eq:wtot}
\end{equation}

The Friedmann constraint reduces the system to four independent variables, which we choose as $(v,y,\mathfrak{h},r)$. With $N\equiv\ln(a/a_0)$, the equations of motion can be written as \footnote{We note that, for a single pressureless component ($r=0$), we recover the equations found in \cite{Bouhmadi-Lopez:2016dzw}.}
\begin{align}
v'&=3\frac{1-v^2}{1+v^2}\left[y\left(1-v^2\right)-v\right],
\label{eq:vprime}\\[4pt]
y'&=\frac{3}{2}\left[y\left(1-y^2+\frac{1}{3}r^2\right)-\frac{1-\mathfrak{h}}{\mathfrak{h}}\bar{V}\exp\left[-\frac{\xi}{9}\left(\frac{v}{1-v^2}\right)^2\right]\left[y-\frac{2\xi}{9}\frac{v\left(1-v^2-vy\right)}{\left(1-v^2\right)^2}\right]\right],
\label{eq:yprime_compact}\\[4pt]
\mathfrak{h}'&=-3(1-\mathfrak{h})\left[\mathfrak{h}\left(1-y^2+\frac{1}{3}r^2\right)-(1-\mathfrak{h})\bar{V}\exp\left[-\frac{\xi}{9}\left(\frac{v}{1-v^2}\right)^2\right]\left[1+\frac{2\xi}{9}\frac{v^2}{\left(1-v^2\right)^2}\right]\right],
\label{eq:hprime}\\[4pt]
r'&=\frac{3}{2}r\left[-\frac{1}{3}-y^2+\frac{1}{3}r^2-\frac{1-\mathfrak{h}}{\mathfrak{h}}\bar{V}\exp\left[-\frac{\xi}{9}\left(\frac{v}{1-v^2}\right)^2\right]\left[1+\frac{2\xi}{9}\frac{v^2}{\left(1-v^2\right)^2}\right]\right],
\label{eq:rprime}
\end{align}
where a prime denotes a derivative with respect to $N$. Eqs.~\eqref{eq:vprime}–\eqref{eq:rprime} together with the definitions above define a closed, bounded dynamical system for the Gaussian 3-form model. 

\begin{table}[t]
    \centering
    \renewcommand{\arraystretch}{1.25} 
    \begin{tabular}{||c|c|c|c|c|c|c|c||}
    \hline
       Point & $v$ & $y$ & $\mathfrak{h}$ & $r$ & $s$ & $w_{\mathrm{tot}}$ & Stability \\
    \hline\hline
       $A_1$ & $0$ & $0$ & $1$ & $1$ & $0$ & $1/3$ & Saddle \\
       $A_2$ & $0$ & $0$ & $1$ & $0$ & $1$ & $0$ & Saddle \\
       $B^{\pm}$ & $\pm\frac{\sqrt{5}-1}{2}$ & $\pm1$ & $1$ & $0$ & $0$ & $-1$ & Attractor \\
       $C$ & $0$ & $0$ & $\frac{\bar{V}}{1+\bar{V}}$ & $0$ & $0$ & $-1$ & Saddle \\
       $D_1^{\pm}$ & $\pm1$ & $0$ & $1$ & $1$ & $0$ & $1/3$ & Repulsive \\
       $D_2^{\pm}$ & $\pm1$ & $0$ & $1$ & $0$ & $1$ & $0$ & Saddle \\
       $E_+^{\pm}$ & $\pm1$ & $1$ & $\mathfrak{h}_{\star}$ & $0$ & $0$ & $-1$ & Saddle \\
       $E_-^{\pm}$ & $\pm1$ & $-1$ & $\mathfrak{h}_{\star}$ & $0$ & $0$ & $-1$ & Saddle \\
    \hline
    \end{tabular}
    \caption{\label{table:fixedpoints3form} 
    Fixed points of the Gaussian 3-form system \eqref{eq:vprime}–\eqref{eq:rprime}. 
    The matter variable $s$ is not an independent degree of freedom but is included here for completeness, being determined by the Friedmann constraint. The quantities marked with $\star$ denote arbitrary values allowed by the constraints. 
    The column $w_{\rm tot}$ shows the value of the total EoS at each point.}
\end{table}

We now classify the fixed points of the system, obtained by solving $v'=y'=\mathfrak{h}'=r'=0$. The results are summarised in Table \ref{table:fixedpoints3form}. We find the same set of fixed points as in \cite{Bouhmadi-Lopez:2016dzw}, but now with two additional points that correspond to radiation domination, $A_1$ and $D_1^{\pm}$. If we eliminate the radiation component, the system reduces to the same table of fixed points reported in \cite{Bouhmadi-Lopez:2016dzw}. Points $A_1$ and $A_2$ correspond to radiation and matter domination, respectively, and both behave as saddle points. Point $B^{\pm}$ corresponds to Little Sibling of the Big Rip\footnote{The LSBR event is considerably milder than the Big Rip singularity. At the LSBR, both the Hubble rate and the scale factor diverge, while the cosmic-time derivative of the Hubble rate remains finite. This abrupt behaviour occurs at infinite cosmic time, where the scalar curvature also diverges. The LSBR is neither a cosmological nor a space–time singularity; however, as it is approached, all structures in the Universe are ultimately torn apart \cite{Bouhmadi-Lopez:2014cca,Albarran:2015cda}.} (LSBR), as previously found in \cite{Morais:2016bev}; these describe for a Gaussian potential the asymptotic behaviour of the 3-form since they act as attractors.

Point $C$ represents a de Sitter state but only appears as a saddle. Points $D_1^{\pm}$ and $D_2^{\pm}$ correspond again to radiation and matter domination; we find that $D_1^{\pm}$ behaves as a repulsor, while $D_2^{\pm}$ is a saddle point. When radiation is switched off, $D_1^{\pm}$ disappears and $D_2^{\pm}$ becomes the repulsor. Finally, the $E_+^{\pm}$ and $E_-^{\pm}$ branches both correspond to de Sitter states; however, they behave as saddle points. Stability can be established through linear perturbation theory in all cases except for $B^{\pm}$, which are non-hyperbolic due to a vanishing eigenvalue of the Jacobian. A centre manifold analysis (presented in Appendix~\ref{app:CM-ours}) shows in detail that $B^{\pm}$ are attractors, confirming that the LSBR solution indeed governs the asymptotic behaviour of the 3-form dynamics.

To illustrate the dynamics of the Gaussian 3-form model, we consider two qualitatively distinct trajectories of the system. 
Both originate near the radiation-dominated repulsor $D_1^{+}$ and terminate at the LSBR attractor $B^{+}$, but they differ in the way matter domination is realised.

\begin{figure*}[t]
  \centering

  \subfigure[Compact variables for the small-field branch.]{
    \includegraphics[width=0.48\textwidth]{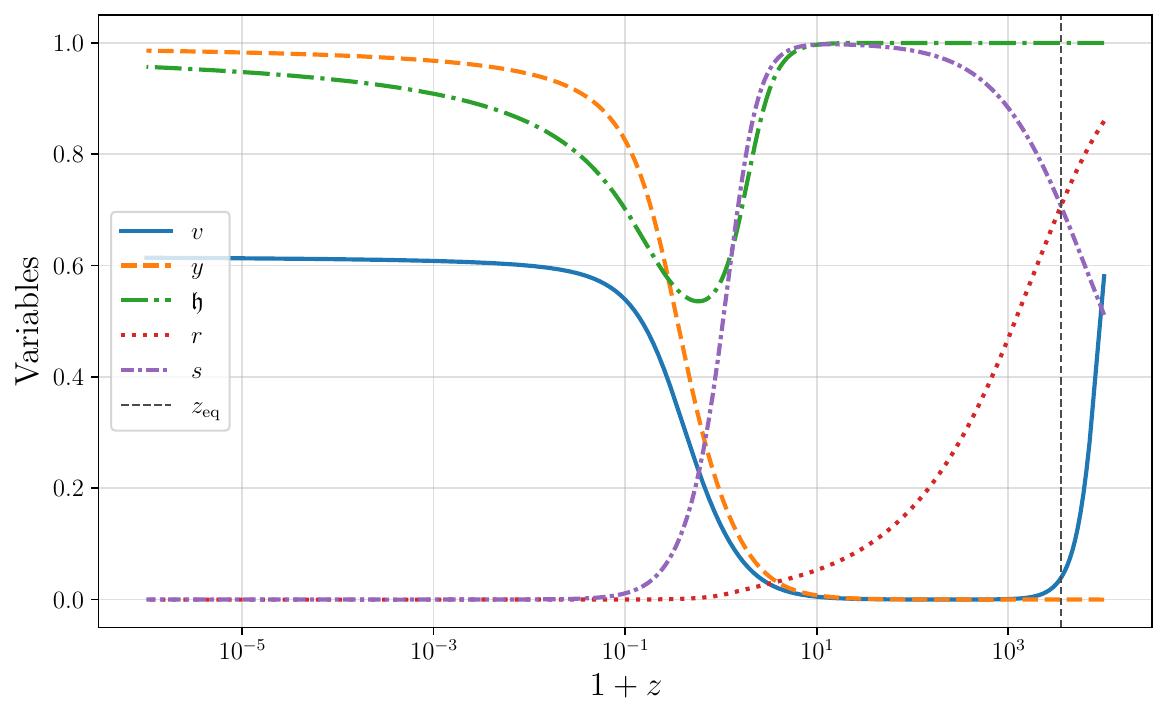}%
    \label{fig:3form-small-vars}
  }\hfill
  \subfigure[EoS for the small-field branch.]{
    \includegraphics[width=0.48\textwidth]{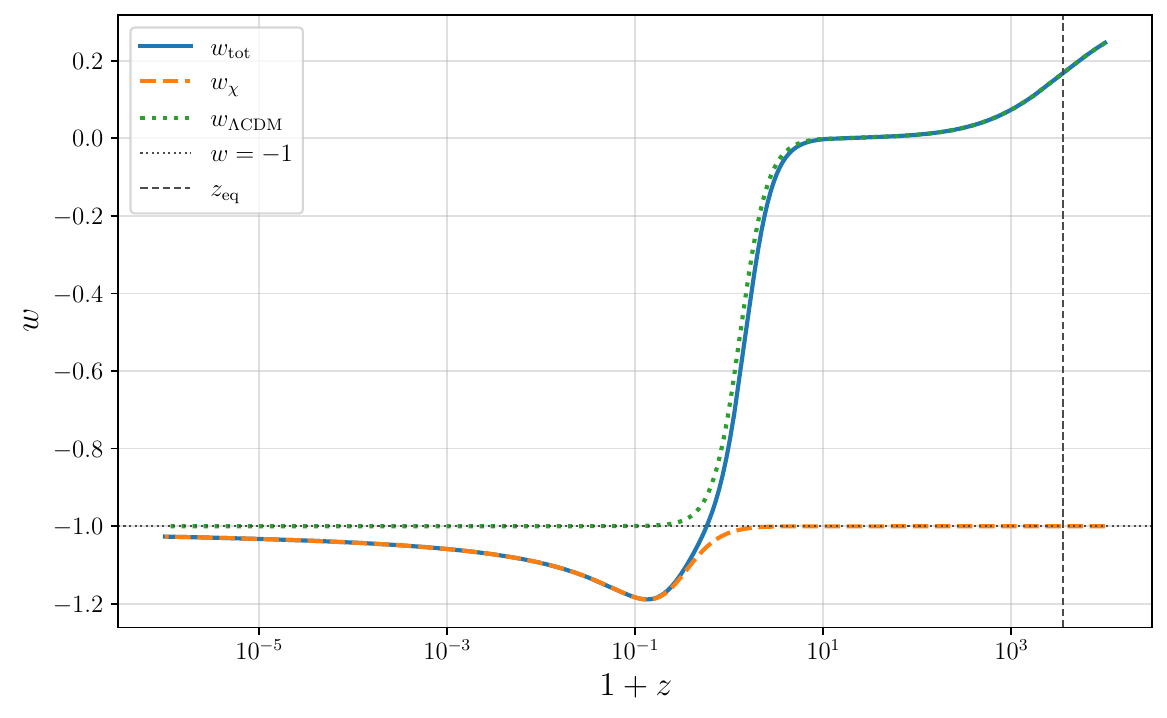}%
    \label{fig:3form-small-w}
  }\\[4pt]

  \subfigure[Compact variables for the large-field branch.]{
    \includegraphics[width=0.48\textwidth]{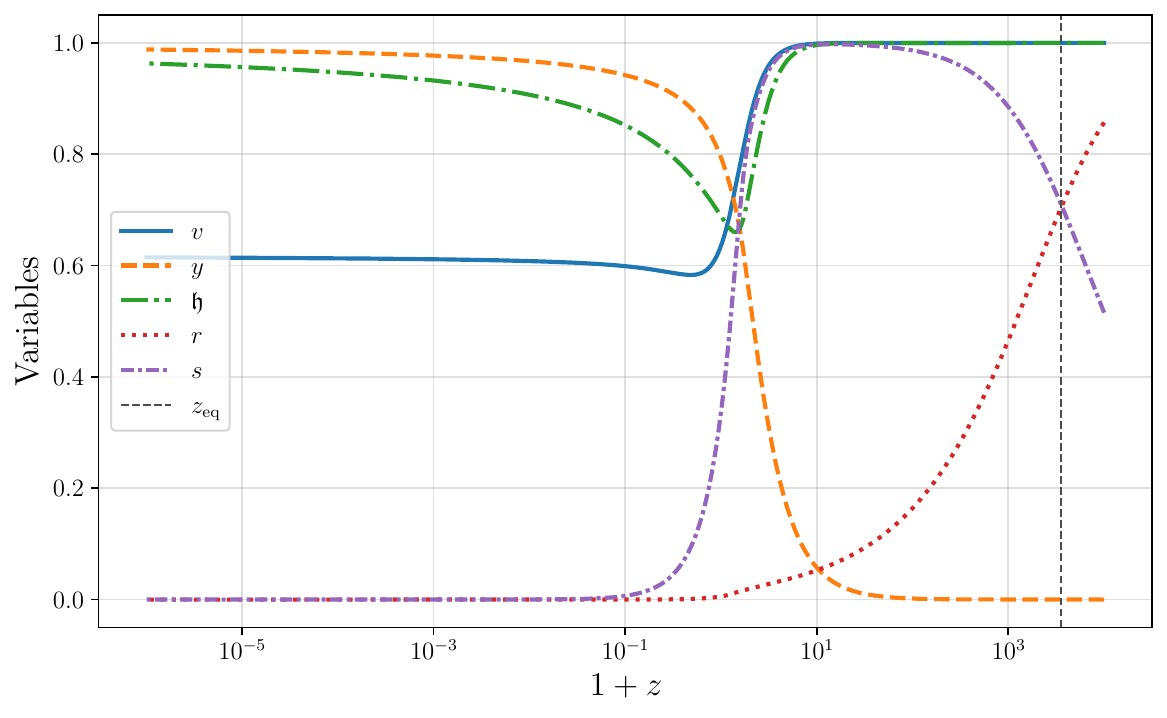}%
    \label{fig:3form-large-vars}
  }\hfill
  \subfigure[EoS for the large-field branch.]{
    \includegraphics[width=0.48\textwidth]{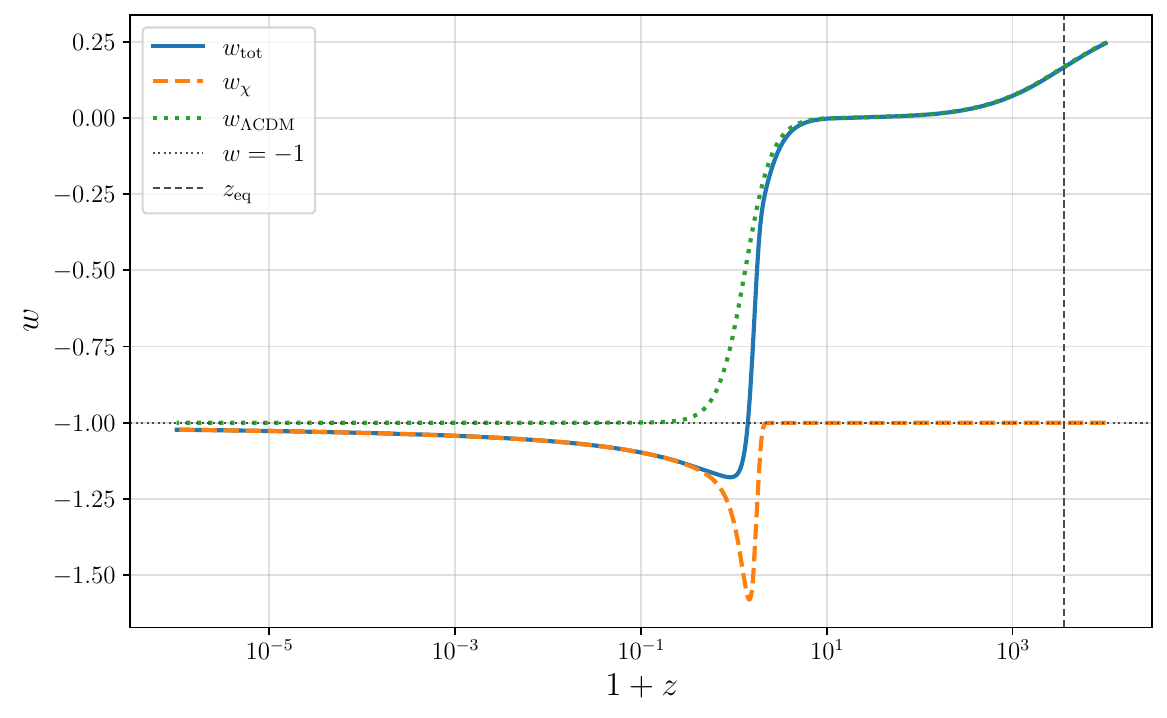}%
    \label{fig:3form-large-w}
  }

  \caption{Representative trajectories of the Gaussian 3-form system, chosen to illustrate how the model can mimic $\Lambda$CDM at early-time while deviating at late-time. 
We set $\bar V=1$ and $\xi\simeq 4.48$, and choose initial conditions such that the redshift of matter–radiation equality \protect\footnotemark\:matches that of $\Lambda$CDM. 
\textbf{Top row:} Evolution of the compact variables $(v,y,\mathfrak{h},r,s)$ and of the two equations of state, $w_{\rm tot}$ and $w_{\chi}$, for a \emph{small-field} trajectory that passes close to the matter saddle $A_2$ (branch $D_1^{+}\!\to A_2 \to B^{+}$). 
\textbf{Bottom row:} Same quantities for a \emph{large-field} trajectory passing close to the matter saddle $D_2^{+}$ (branch $D_1^{+}\!\to D_2^{+} \to B^{+}$). 
In both cases, the 3-form dynamics eventually drive the system away from matter domination and towards the LSBR attractor $B^{+}$, producing a late-time phantom crossing ($w_{\rm tot}< -1$) and an asymptotic approach to the LSBR regime with $w_{\rm tot}\to -1^{-}$.}
  \label{fig:3form-trajectory-multibranch}
\end{figure*}

\paragraph{Small-field branch: $D_1^{+}\!\to A_2\to B^{+}$.}
In the first trajectory, shown in the top row of Fig.~\ref{fig:3form-trajectory-multibranch}, the 3-form field remains close to the top of the potential around the onset of matter domination.  
The system evolves from $D_1^{+}$ towards the matter saddle $A_2$, characterised by 
$s\simeq 1$, $r\simeq 0$, and a small field amplitude $v\approx 0$.  
However, remaining near $A_2$ for a prolonged period requires the field to lie extremely close to the maximum of the Gaussian potential, where $V_{,\chi}\approx 0$.  
Because the potential steepens rapidly away from $\chi=0$, even tiny deviations in the initial conditions cause the field to roll away from the hilltop.  
Once the field departs from the vicinity of $A_2$, the evolution progressively approaches the LSBR attractor $B^{+}$, which controls the remote future dynamics of the system.
\REV{As the field lies near the top of the potential throughout most of the matter-dominated era, the 3 form is in practice a cosmological constant until recently, and thus it is a reasonable assumption that any mode re-entered after the matter-radiation equality is completely stable.}

\paragraph{Large-field branch: $D_1^{+}\!\to D_2^{+}\to B^{+}$.}
The second trajectory, shown in the bottom row of Fig.~\ref{fig:3form-trajectory-multibranch}, evolves from $D_1^{+}$ towards the matter saddle $D_2^{+}$.  
This point also describes matter domination ($s\simeq 1$, $w_{\rm tot}\simeq 0$), but with a large field amplitude $v\approx 1$, so the 3-form is already displaced from the top of the potential at the beginning of the matter era.  
In contrast to the previous case, this branch does \emph{not} require fine-tuning: a wide range of initial conditions naturally drives the evolution toward $D_2^{+}$.  
After the matter phase, the system again converges smoothly towards the LSBR attractor $B^{+}$, revealing that the LSBR regime constitutes a robust far-future behaviour of the Gaussian 3-form model. This behaviour is also consistent with our observational fits, which favour the large-field branch over the fine-tuned small-field trajectory (see Section \ref{sec:fit}).
\REV{The price to pay is of course the stability of the 3-form perturbation, as the 3-form remains far away from the top of the potential for the majority of the matter-dominated era. With the benefit of the hindsight from section~\ref{sec:evolution}, the system is dominated by the Hubble flow with $a^3 \chi \sim$ constant, leading to an extreme runaway scenario with a growth factor of $O (a_{\rm re}^{-2})$ where $a_{\rm re}$ is the scale factor at which the mode re-entered. Therefore, one should not interpret the 3-form as a valid perturbative description during the matter-dominated era in this branch.}

\paragraph{Equation of state.}
The right panels of Fig.~\ref{fig:3form-trajectory-multibranch} show the corresponding evolution of both the total equation of state, $w_{\rm tot}$, and the 3-form equation of state, $w_{\chi}$.  
\REV{Despite vastly different behaviour at the background and the perturbative level, i}\COMREV{I}n both trajectories, the standard early-time sequence is recovered: radiation domination ($w_{\rm tot}=1/3$), followed by matter domination ($w_{\rm tot}=0$)\REV{, with the 3-form as an effective description that cannot be applied at the perturbative level}.  
At late-time, the 3-form naturally drives the Universe into a phantom regime, $w_{\rm tot}< -1$, as expected from the negative slope of the Gaussian potential; correspondingly, $w_{\chi}$ also evolves into the phantom domain and dominates the total EoS once the field becomes the leading component.\REV{~As the 3-form approaches the top of the potential, the perturbation stabilises, leading to a consistent description of the perturbation evolution. }  
In the distant future, the system approaches the LSBR regime, characterised by $w_{\rm tot}\to -1^{-}$ and an ever-increasing Hubble rate that diverges only at infinite cosmic time.  
Although the LSBR is reached asymptotically, it still represents an abrupt future event in the sense that all bound structures are ultimately disrupted as the Hubble rate grows without bound.  
Nevertheless, this behaviour emerges only at extremely late-time, and both branches follow distinct dynamical paths before settling into the same asymptotic LSBR attractor.

\footnotetext{In the $\Lambda$CDM model we compute the redshift of matter–radiation equality using 
\[
z_{\mathrm{eq}} = 2.5 \times 10^4\, \Omega_{\rm m0} h^2 \left(\tfrac{T_{\mathrm{CMB}}}{2.7}\right)^{-4} \quad \text{\cite{Eisenstein:1997ik}},
\] 
with $\Omega_{\rm m0}=0.3$, $h=H_0/100=0.7$, and $T_{\mathrm{CMB}}=2.7255\,\mathrm{K}$ \cite{Fixsen:2009ug}. 
We then set $\Omega_{\rm r0}=\Omega_{\rm m0}/(1+z_{\mathrm{eq}})$, while the dark energy density parameter follows from the Friedmann constraint in $\Lambda$CDM, $\Omega_{\Lambda0}=1-\Omega_{\rm m0}-\Omega_{\rm r0}$. These values are used to perform the $\Lambda$CDM numerical calculations shown in the figures.}


\section{Observational fit}
\label{sec:fit}

We utilise Cobaya's Monte-Carlo-Markov-Chain (MCMC) sampler \cite{Torrado:2020dgo, Lewis:2002ah, Lewis:2013hha, 2005math......2099N} and CAMB Boltzmann solver \cite{Lewis:1999bs,Howlett:2012mh} to generate the posterior distribution of the full cosmological parameters, including parameters for early-time CMB physics, i.e., $\Omega_{\rm b0} h^2$, $\Omega_{\rm c0} h^2$, $\theta_{\rm MC}$, $\ln A_s$, $n_s$ and $\tau_{\rm reio}$\footnote{ `The parameter $H_0 = 100 h$ is the Hubble constant derived from the characteristic acoustic peak angular diameter of CMB $\theta_{\rm MC}$ assuming a cosmic evolution where $H_0$ can be factored out as a scale \cite{Hu:1995en}. The model we consider adheres to this assumption. $\Omega_{\rm b0}$ and $\Omega_{\rm c0}$ are the density parameter of the baryon and the cold dark matter at the present time. We assume the minimal model of 3 neutrino generations with 1 massive neutrino of mass $0.06$eV and an effective number of $3.044$. The dark energy density $\Omega_{\Lambda0}$ can be derived following these assumptions. Following standard practice, the primordial perturbation spectrum is red-tilted with the spectrum power $A_s$ and the spectrum tilt $1-n_s$. The reionization follows the standard $\tanh$ form with the optical depth $\tau_{\rm reio}$.}, for datasets that will be introduced in section~\ref{sec:data}. For each dataset, $3 \sim 6$ chains are deployed based on the early convergence test. Slow, non-convergent chains are dropped, leaving $2 \sim 3$ chains for statistical analysis. We utilise the stopping criteria of Gelman-Rubin\footnote{The parameter R is defined as $R = \underset{\theta}{\max} \sqrt{1 + \left( \left< \left< \theta \right>_{\rm inner}^2 \right> - \left< \left< \theta \right>_{\rm inner} \right>^2 \right) \Big{/} \left< \left<\theta^2\right>_{\rm inner} - \left< \theta \right>_{\rm inner}^2 \right>}$, where $\theta$ are the parameters, $\left<\dots\right>_{\rm inner}$ is the average over an entire chain for inter-chain $R$, or over a segment of a chain for intra-chain $R$ with each segment $40N_\theta$ where $N_\theta$ is the number of parameters, and $\left<\dots\right>$ without subscript is the average over chains for inter-chain $R$ or over segments within a single chain for intra-chain $R$.}, with $R-1 < 0.1$ intra-chain and $R-1 < 0.01$ inter-chain.

To compare between different datasets and models, we primarily utilise 6 probes. The traditional Bayesian evidence $B$, deviance information criteria (DIC), and widely applicable information criteria (WAIC) \cite{10.5555/2567709.2502609} are for model comparison, while the Bayesian ratio $R$, the goodness of fit (GoF), and suspiciousness $S$ \cite{DES:2020hen, Raveri:2019gdp, PhysRevD.100.023512} are for data tension detection. For detailed explanation of each probe please refer to appendix~\ref{sec:stat}. To interpret the probes, all we need to know is that, except for the goodness of fit and suspiciousness, they follow Jefferys' scale, with larger value meaning statistically disfavoured and vice versa. For the goodness of fit and suspiciousness, we convert them into traditional $\sigma$-values.

Regarding the prior, we follow the standard practice of \cite{Planck2018} for the early-time cosmological parameters, as shown in table~\ref{table:prior}. Additional BBN (big bang nucleosynthesis) prior on the baryon density $\Omega_{\rm b0} h^2$ and a cut $40 < H_0 < 100$ on the Hubble constant are imposed across all models and datasets considered in this work to stabilise chains without CMB data. For late-time physics, i.e., the dark energy model, we put the 3-form dark energy model in section~\ref{sec:model} with the Gaussian potential against the $\Lambda$CDM model.

Furthermore, a spatially flat FLRW universe is assumed, leading to a 6-parameter model for $\Lambda$CDM. For the 3-form model, the traditional approach of setting $V_\ast \sim \rho_{\rm DE,0} \equiv 3 H^2 k^{-2} \Omega_{\Lambda0}$ is not suitable as $\chi = 0$ is the maximum of the potential rather than the minimum. Interestingly, empirically $\mathrm{KE}_i + V_\ast$ appears to be a good proxy for the current dark energy density as shown in table~\ref{table:result_scalar}, with $\mathrm{KE}_i = \left(\dot\chi_i + 3 H_i \chi_i \right)^2/2$ the initial kinetic energy of 3-form field. This is most likely the consequence of 3-form tending toward a cosmological constant before the dark-energy-dominated era according to the observation, leading to a suppression factor of $a^3$ for the effect of potential on the evolution of the kinetic energy. However, 3-form does occasionally climb up or roll down the potential at low-$z$, resulting in a loss/gain of the kinetic energy much greater than $O(0.1\%)$ of the total energy density, i.e., the uncertainty we expect from the acoustic peak scale. Therefore, one should not impose the constraint of $\mathrm{KE}_i + V_\ast = \rho_{\rm DE,0}$ either. Instead, we impose a Gaussian likelihood of width $0.001/\sqrt{2}$ on $\mathcal{E}_{H_0} \equiv 1 - H_{0,{\rm EOM}}^2/ H_0^2$ reminiscent of an ad hoc Gaussian constraint, where $H_{0,{\rm EOM}}$ is the Hubble constant derived from the equation of motion and $H_0$ is the Hubble constant derived from $\theta_{\rm MC}$, to maintain a consistent background equation of motion during the forward-time integration of CAMB.\footnote{Even though this constraint is tautological in backward-time integration where $\rho_{\rm DE,0}$ can be determined analytically, we stick to the forward-time integration to implement the prior near the radiation-dominated fixed point $D^+$ with a positive definite initial field strength $\chi$. This gives us a better control on the evolution of 3-form and thus interpretability of the model parameters, at the cost of an additional parameter.} 
We will subtract the likelihood of the constraint from all the likelihoods presented in this work, thus converting it into a prior probability\footnote{The associated prior volume for the Bayesian evidence in Eq.~\eqref{eq:BE} can be derived from the Bayesian ratio in the null hypothesis test.} of $\mathcal{E}_{H_0}$. With additional action parameters of $V_\ast / \rho_{\rm DE,0}$, $\log_{10} \xi$ and the initial conditions of $\log_{10} (a_i^3 \sqrt\xi \kappa \chi_i)$ and $\sqrt{3} \kappa H_i \chi_c y_i / (70{\rm km/s/Mpc}) = \left( \dot\chi_i + 3 H_i \chi_i \right) / \rho_{{\rm crit},h=0.7}^{1/2}$ at $a_i/a_0 = 10^{-6}$, we may consider the 3-form dark energy as a (6+4)-parameter model, with one strongly constraint direction along $\mathcal{E}_{H_0}$.

The prior for the 3-form dark energy model is chosen given $\rho_{\rm DE,0} \approx \mathrm{KE}_i + V_\ast$ and the dynamical variables of $V$ and $F_{0ijk} F^{0ijk} = \left(\dot\chi + 3 H \chi \right)^2$. To match with the Gaussian potential, we choose a logarithmic prior on $a_i^3 \sqrt\xi \kappa \chi_i$ and $\xi$, given us direct control of when the 3-form field climbs up the potential (and when the speed of sound becomes positive).\footnote{Notice that the prior on $\dot\chi_i + 3 H_i \chi_i$ does cover the fixed point $E$ (kinetic-energy-dominated). However, it is strongly suppressed by the flat prior on $V_\ast$ and $\dot\chi_i + 3 H_i \chi_i$. This is a deliberate choice to prevent MCMC from becoming trapped around $E$ which is indistinguishable from the $\Lambda$CDM model. On the other hand, we explicitly exclude fixed points $A_1$ and $A_2$, as the \emph{small-field branch} requires significant fine-tuning.} \COMREV{Furthermore, we verify that the instability associated with a negative speed of sound at the early time does not materialise by deploying}\REV{Finally, as discussed in the previous section, the 3-form dark energy with a Gaussian potential is inherently unstable in the early universe, and one should consider it as an effective description of some other underlying field. For the numerical analysis, we deploy} \COMREV{two}\REV{three} different perturbation code\REV{s} within CAMB\REV{~that does not surfer from the gradient instability}, \COMREV{one}\REV{two} with \COMREV{a }modified scalar perturbation evolution governing the physical degrees of freedom of the 3-form field $\delta\chi$\REV{\footnote{\REV{These two perturbation codes differ by a scaling factor in the definition of the perturbed conjugate momentum $\delta\pi$, with $\delta\pi = \delta \dot\chi$ in the first realisation and $\delta\pi = \delta (\dot\chi + 3 H\chi)$ in the second\HWC{, where we assume the perturbation of the form of a scalar field $\chi$ minimally coupled to gravity}.}
}}, and another with the dark energy perturbation turned off. These two realisations produce identical CMB power spectrum. In the rest of this work we will keep the scalar dark energy perturbation.

\begin{table}[]
\centering
\begin{tabular}{|c|c|c c c|}
\hline
\multirow{6}{*}{$\Lambda$CDM parameters}
& $\Omega_{\rm b0} h^2$             & norm  & $0.0222$  & $0.0005$  \\
& $\Omega_{\rm c0} h^2$             & flat  & $0.001$   & $0.99$    \\
& $100\theta_{\rm MC}$              & flat  & $0.5$     & $10$      \\
& $\ln ( 10^{10} A_s )$             & flat  & $1.61$    & $3.91$    \\
& $n_s$                             & flat  & $0.8$     & $1.2$     \\
& $\tau_{\rm reio}$                 & flat  & $0.01$    & $0.8$     \\
\hline
\multirow{4}{*}{Extended parameters}
& $V_\ast/\rho_{\rm DE,0}$          & flat  & $0$       & $2$       \\
& $\log_{10} \xi$                   & flat  & $-7$      & $\log_{10}(9/2)$     \\
& $\log_{10} (a_i^3 \sqrt\xi \kappa \chi_i)$& flat  & $-3.5$    & $0.5$     \\
& $(\dot\chi_i +3H_i \chi_i) \rho_{{\rm crit,}h=0.7}^{-1/2}$& flat  & $-5/4$    & $5/4$     \\
\hline
\multirow{2}{*}{Constraint}
& $\mathcal{E}_{H_0}$               & norm  & $0$       & $0.001/\sqrt{2}$  \\
& $H_0$                             & flat  & $40$      & $100$     \\
\hline
\end{tabular}
\caption{\label{table:prior}
The model parameters of $\Lambda$CDM model and the 3-form dark energy model considered in this work. For the normal distribution two columns provide the mean and the width respectively, and for the flat distribution two columns corresponds to the minimum and the maximum. }
\end{table}

This section is divided into 3 subsections. Section \ref{sec:data} explains the dataset considered in this paper. Section \ref{sec:result} presents the main result of this work, i.e, the cosmological parameters of the 3-form dark energy model under various datasets. Section \ref{sec:evolution} discusses the effect of the 3-form dark energy on the cosmic evolution.

\subsection{Data}
\label{sec:data}

We consider multiple astrophysical and cosmological datasets. For simplicity, they are grouped together according to the theoretical origin as CMB (\textit{Planck} 2018 low-$\ell$ TTEE \cite{Planck:2019nip}, NPIPE CamSpec high-$\ell$ TTTEEE \cite{Rosenberg:2022sdy}, PR4 CMB lensing \cite{Carron:2022eyg, Carron:2022eum}) , BAO (DESI DR1 without full-shape \cite{DESI:2024mwx, DESI:2024lzq, DESI:2024uvr}), SNe (Pantheon+ SNe catalogue \cite{Brout:2022vxf} without anchoring of SNe standardised absolute magnitude)\footnote{The low-$z$ dataset is separated from the SNe catalogue in our analysis, reducing the tension from $>5\sigma$ for the combined Patheon+SH0ES to $4.3\sigma$. We do so for the orthogonality between Pantheon+ and low-$z$ dataset, which turns out to be irrelevant as low-$z$ anchors dominates in $H_0$ tension.}, low-$z$ (Riess et al. low-$z$ anchors \cite{Riess:2020fzl}), and DES Y1 morphological datasets \cite{DES:2017myr}. Additional $f\sigma_8$ dataset (table 2 of \cite{Avila:2022xad}, with data entries from \cite{eBOSS:2020lta, Turnbull:2011ty, Achitouv:2016mbn, Beutler:2012px, Feix:2015dla, BOSS:2016wmc, BOSS:2013eso, Blake:2012pj, Nadathur:2019mct, BOSS:2013mwe, Wilson:2016ggz, eBOSS:2018yfg, Okumura:2015lvp}), SDSS DR7 LRG dataset \cite{Reid:2009xm} and eBOSS DR14 Ly-$\alpha$ dataset \cite{eBOSS:2018qyj} are presented in our figures in section~\ref{sec:evolution}, but are excluded from our cosmological fit.

We do not consider the distance priors of CMB \cite{Chen:2018dbv}, and instead rely on CAMB to capture as much information as possible. This is crucial for the verification of the marginality of the dark energy perturbation. While in the end we do obtain identical results \COMREV{with or without}\REV{for both realisations of} the dark energy perturbation, this is a priori unknown to us. With Planck as our base dataset, we then progressively add in later-time datasets in the following order: BAO, SNe, low-$z$, and DES Y1. For the evaluation of the Bayesian ratio and suspiciousness between each dataset, we also consider 3 smaller datasets: BAO, SNe+low-$z$ and DES Y1.

Due to time constraint, even though the Bayesian evidence, DIC, WAIC, etc. have stabilised, we cannot meet the stopping criteria when fitting DES Y1 dataset with the 3-form dark energy model (Inter-chain $R-1 \sim 0.08$ for $H_0$ as the model often sticks around the initial point. Of 6 chains we perform only 3 successfully explore a large parameter space.), and one should be cautious against reading too much into its posterior distribution. 

\subsection{Cosmological parameters}
\label{sec:result}

For the visualisation of the cosmological fit, we utilise GetDist package \cite{Lewis:2019xzd} and choose a different 
set of parameters from the ones in table~\ref{table:prior}. For the early-time CMB physics we stick with the well-accepted 6-dimensional parameter space of $\Omega_{\rm b0} h^2$, $\Omega_{\rm c0} h^2$, $100\theta_{\rm MC}$, $\ln (10^{10} A_s)$, $n_s$ and $\tau_{\rm reio}$. For the late-time physics, we choose the standard observables of Hubble constant $H_0$, total matter density parameter at current time $\Omega_{\rm m0}$, the matter spectrum power $\sigma_8$ at the scale of $8{\rm Mpc}/h$, and the rescaled matter spectrum power $S_8 = \sigma_8\sqrt{\Omega_{\rm m0} / 0.3}$. For the model parameters of the 3-form dark energy, we choose $(\dot\chi_i + 3 H_i \chi_i ) / \rho_{\rm DE,0}^{1/2}$ and $(\mathrm{KE}_i + V_\ast ) / \rho_{\rm DE,0}$ instead of $(\dot\chi_i + 3 H_i \chi_i ) / \rho_{{\rm crit}, h = 0.7}^{1/2}$ and $\log_{10} (V_\ast / \rho_{\rm DE,0})$ to better match with the total energy density fraction of the 3-form.

With our choice of the parameters fixed, let us first analyse if the 3-form dark energy model could affect the early-time CMB physics, thus leading to different late-time predictions that may ease some cosmological tension. To achieve this goal, we consider a series of MCMC analyses that gradually includes CMB, BAO, SNe, low-$z$ and DES Y1 datasets.

\begin{table}
\centering
\scriptsize
\begin{tabular}{|c||c|c|c|c|c|}
\hline
                        & CMB               & Left + BAO            & Left + SNe            & Left + low-$z$        & Left + DES Y1 \\
\hhline{|=#=|=|=|=|=|}
$10^3\Omega_{\rm b0}h^2$& $22.19    \pm0.13$& $22.29    \pm0.12$    & $22.26    \pm0.12$    & $22.34    \pm0.12$    & $22.38    \pm0.12$    \\
$10^3\Omega_{\rm c0}h^2$& $119.7    \pm1.0$ & $118.26   \pm0.81$    & $118.56   \pm0.79$    & $117.80   \pm0.76$    & $117.27   \pm0.72$    \\
$10^5\theta_{\rm MC}$   & $1040.77  \pm0.25$& $1040.94  \pm0.24$    & $1040.91  \pm0.24$    & $1041.01  \pm0.24$    & $1041.05  \pm0.24$    \\
$\ln (10^{10} A_s)$ & $3.037    \pm0.014$   & $3.044    \pm0.014$   & $3.043    \pm0.014$   & $3.047    \pm0.014$   & $3.046    \pm 0.014$  \\
$n_s$               & $0.9636   \pm0.0040$  & $0.9672   \pm0.0036$  & $0.9665   \pm0.0036$  & $0.9685   \pm0.0036$  & $0.9693   \pm0.0036$  \\
$\tau_{\rm reio}$   & $0.0524   \pm0.0071$  & $0.0571   \pm0.0071$  & $0.0562   \pm0.0070$  & $0.0587   \pm0.0071$  & $0.0589   \pm0.0071$  \\
\hline
$H_0$               & $67.24    \pm0.46$    & $67.89    \pm0.36$    & $67.76    \pm0.35$    & $68.12    \pm0.34$    & $68.36    \pm0.32$    \\
$\Omega_{\rm m0}$   & $0.3154   \pm0.0064$  & $0.3064   \pm0.0048$  & $0.3082   \pm0.0047$  & $0.3034   \pm0.0044$  & $0.3003   \pm0.0042$  \\
$\sigma_8$          & $0.8077   \pm0.0055$  & $0.8064   \pm0.0056$  & $0.8067   \pm0.0056$  & $0.8061   \pm0.0057$  & $0.8041   \pm0.0055$  \\
$S_8$               & $0.828    \pm0.011$   & $0.8149   \pm0.0090$  & $0.8176   \pm0.0090$  & $0.8107   \pm0.0087$  & $0.8044   \pm0.0080$  \\
\hline
DIC                 & $5497.90  \pm0.12$    & $5507.21  \pm0.37$    & $6209.34  \pm0.14$    & $6217.77  \pm0.43$    & $6477.38  \pm0.28$    \\
WAIC                & $5499.09  \pm0.50$    & $5507.85  \pm0.18$    & $6210.36  \pm0.21$    & $6218.68  \pm0.21$    & $6481.16  \pm0.22$    \\
$-\ln B$            & $5499.3   \pm1.1$     & $5508.13  \pm0.73$    & $6210.51  \pm0.44$    & $6219.1   \pm1.4$     & $6479.9   \pm1.2 $    \\
\hline
\end{tabular}
\caption{\label{table:result_LambdaCDM} Mean and standard deviation of cosmological parameters, late-time observables, and statistical probes for $\Lambda$CDM model. From left to right are gradually larger datasets that progressively add in datasets of CMB, BAO, etc., as defined in section~\ref{sec:data}.}
\end{table}

\begin{table}
\centering
\scriptsize
\begin{tabular}{|c||c|c|c|c|c|}
\hline
                        & CMB               & Left + BAO            & Left + SNe            & Left + low-$z$        & Left + DES Y1 \\
\hhline{|=#=|=|=|=|=|}
$10^3\Omega_{\rm b0}h^2$& $22.19\pm0.13$    & $22.25    \pm0.13$    & $22.23    \pm0.12$    & $22.26    \pm0.13$    & $22.30    \pm0.13$    \\
$10^3\Omega_{\rm c0}h^2$& $119.6    \pm1.1$ & $118.79   \pm0.95$    & $118.89   \pm0.86$    & $118.79   \pm0.98$    & $118.3    \pm1.0$     \\
$10^5\theta_{\rm MC}$   & $1040.78  \pm0.26$& $1040.88  \pm0.25$    & $1040.88  \pm0.24$    & $1040.88  \pm0.26$    & $1040.93  \pm0.24$    \\
$\ln (10^{10} A_s)$ & $3.037    \pm0.014$   & $3.041    \pm0.014$   & $3.040    \pm0.014$   & $3.039    \pm0.015$   & $3.038    \pm0.015$   \\
$n_s$               & $0.9640   \pm0.0040$  & $0.9662   \pm0.0037$  & $0.9658   \pm0.0037$  & $0.9663   \pm0.0038$  & $0.9670   \pm0.0039$  \\
$\tau_{\rm reio}$   & $0.0523   \pm0.0070$  & $0.0550   \pm0.0073$  & $0.0547   \pm0.0072$  & $0.0543   \pm0.0077$  & $0.0540   \pm0.0077$  \\
\hline
$\log_{10} \xi$     & $\color{cyan}-3.2$    & $\color{cyan}-3.2$    & $\color{cyan}-3.4$    & $\color{cyan}-3.4$    & $\color{cyan}-3.4$    \\
$({\rm KE}_i+V_\ast)/\rho_{\rm DE,0}$   & $1.0000   \pm{\color{cyan}0.0021}$& $1.0002   \pm{\color{cyan}0.0027}$& $1.0000   \pm{\color{cyan}0.0017}$& $1.0004   \pm{\color{cyan}0.0020}$& $1.0006   \pm{\color{cyan}0.0038}$\\
$\log_{10} (a_i^3 \sqrt\xi \kappa \chi_i)$  &$-2.3^{+0.6}_{\color{white}-?}$&$-1.6^{+0.3}_{\color{white}-?}$& $\color{BrickRed}{\color{black}-1.86}^{+0.69}_{\color{cyan}-0.59}$& $-1.27^{+0.28}_{\color{cyan}-0.21}$ & $-1.23^{+0.27}_{\color{cyan}-0.23}$ \\
$(\dot\chi_i + 3 H_i \chi_i)\rho_{\rm DE,0}^{-1/2}$ & $0.02\pm0.51$ & $0.02\pm0.51$ & $-0.03\pm0.48$& $-0.06\pm0.46$& $-0.08\pm0.45$\\
\hline
$H_0$   & $\color{blue}{\color{black}67.57}^{+0.58}_{\color{black}-0.75}$   & $\color{blue}{\color{black}68.29}^{+0.56}_{\color{black}-0.61}$   & $\color{blue}{\color{black}68.02}^{+0.45}_{\color{black}-0.46}$   & $68.92    \pm0.59$    & $69.21    \pm0.60$    \\
$\Omega_{\rm m0}$& $0.3116^{+0.0090}_{\color{blue}-0.0075}$ & $0.3030   \pm0.0060$  & $0.3062   \pm0.0051$  & $0.2984   \pm0.0052$  & $0.2950   \pm0.0053$  \\
$\sigma_8$  & $\color{blue}{\color{black}0.8076}^{+0.0060}_{\color{black}-0.0063}$  & $0.8080   \pm0.0070$  & $0.8069   \pm0.0065$  & $0.8101   \pm0.0069$  & $0.8083   \pm0.0066$  \\
$S_8$               & $0.822^{+0.013}_{\color{cyan}-0.011}$ & $0.8119   \pm0.0094$  & $0.8151   \pm0.0091$  & $0.8079   \pm0.0086$  & $0.8015   \pm0.0085$  \\
\hline
DIC                 & $5498.72  \pm0.01$    & $5507.62  \pm0.54$    & $6209.59  \pm0.53$    & $6216.09  \pm0.31$    & $6476.4   \pm1.2$     \\
WAIC                & $5499.58  \pm0.21$    & $5508.21  \pm0.29$    & $6210.14  \pm0.45$    & $6217.51  \pm0.65$    & $6479.94  \pm0.06$    \\
$-\ln B$            & $5500.0   \pm1.1$     & $5508.5   \pm1.2$     & $6210.04  \pm0.77$    & $6216.24  \pm0.20$    & $6475.82  \pm0.11$    \\
\hline
$\Delta$DIC         & $   0.82  \pm0.15$    & $   0.40  \pm0.44$    & $   0.25  \pm0.56$    & $  -1.68  \pm0.60$    & $  -1.0   \pm1.3$     \\
$\Delta$WAIC        & $   0.49  \pm0.63$    & $   0.36  \pm0.36$    & $  -0.22  \pm0.52$    & $  -1.17  \pm0.70$    & $  -1.22  \pm0.27$    \\
$-\Delta\ln B$      & $   0.8   \pm1.7$     & $   0.3   \pm1.5$     & $  -0.47  \pm0.94$    & $  -2.9   \pm1.7$     & $  -4.1   \pm1.4$     \\
\hline
\end{tabular}
\caption{\label{table:result_scalar} Mean and standard deviation of cosmological parameters, late-time observables, and statistical probes for the 3-form dark energy model in section~\ref{sec:model}. From left to right are gradually larger datasets of CMB, CMB + BAO, CMB + BAO + SNe, etc., as defined in section~\ref{sec:data}. $\Delta{\rm ICs}$ are with respect to $\Lambda$CDM model presented in table~\ref{table:result_LambdaCDM}. For parameters not following Gaussian distribution we provide the median and 68\% lower and upper bounds (if valid) instead, with colour coding for how heavy the tail is ({\color{BrickRed}red} for short tail, black for Gaussian, {\color{blue}blue} for exponential, and {\color{Cyan}cyan} for long tail.) For single-sided distributions we report the modal and the single-sided 68\% bound instead.}
\end{table}

The mean and the standard deviation of the cosmological parameters, late-time observables, and statistical probes of these analyses are shown in table~\ref{table:result_LambdaCDM} for $\Lambda$CDM model and in table~\ref{table:result_scalar} for the 3-form dark energy model. From the tables, one can infer that the model we consider in this work does provide a statistically marginal benefit once the low-$z$ dataset is included, as most statistical probes weakly but consistently prefer 3-form model over $\Lambda$CDM model. The statistical preference is supported by the increase of the predicted values of $H_0$ by 3-form model relative to $\Lambda$CDM model, suggesting a partial resolution of the Hubble tension.

\begin{figure}
\centering
\includegraphics[width = 0.96\textwidth]{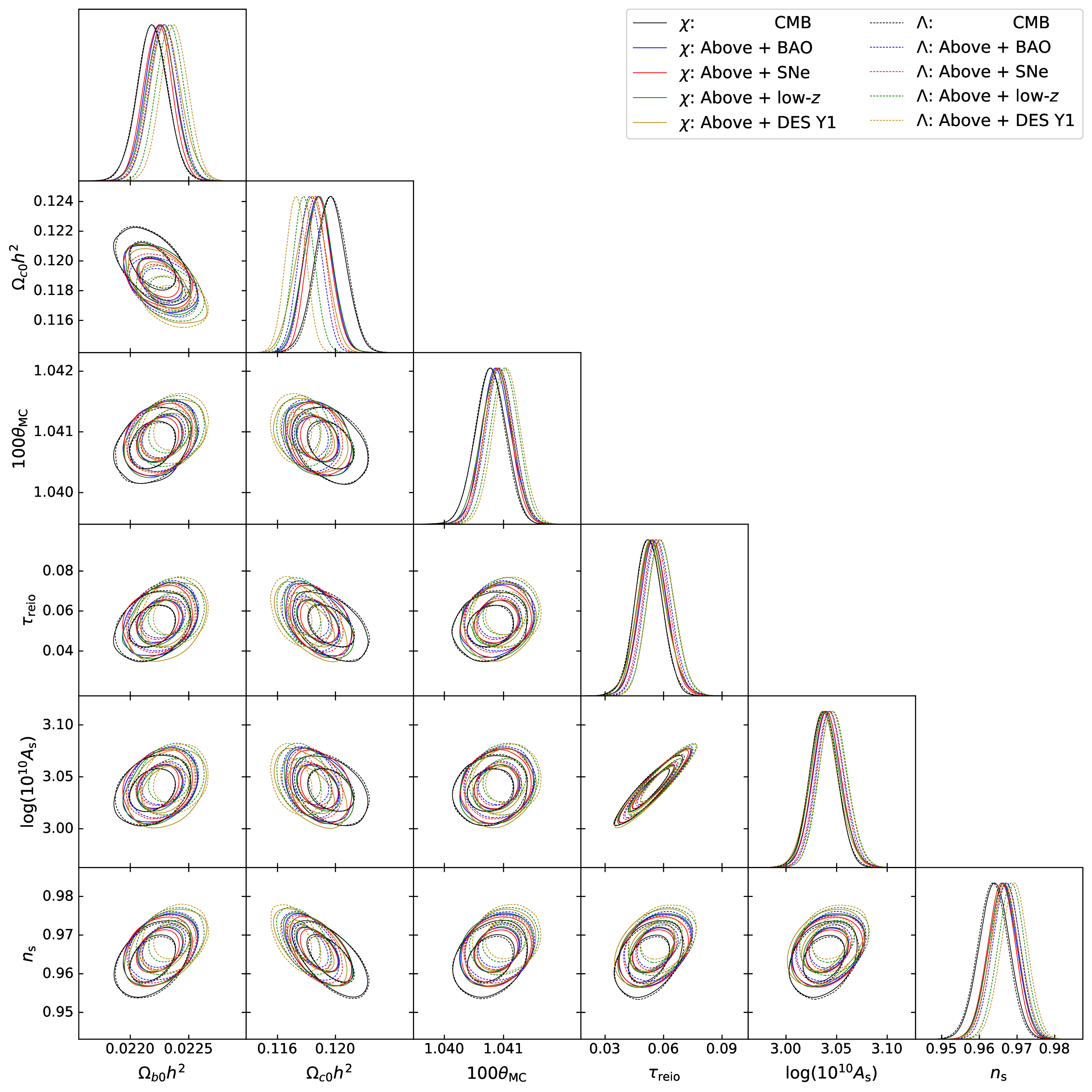}
\caption{\label{fig:fit_CMBplus_early_params} 68\% and 95\% C.L. posterior distribution of the early-time parameters from CMB (black), CMB + BAO ({\color{RoyalBlue}blue}), CMB + BAO + SNe ({\color{BrickRed}red}), CMB + BAO + SNe + low-$z$ ({\color{ForestGreen}green}), and CMB + BAO + SNe + low-$z$ + DES Y1 ({\color{Dandelion}yellow}) data, with $\Lambda$CDM model ($\Lambda$) the dashed contours and the 3-form dark energy model ($\chi$) the solid contours. The result demonstrates the superior stability of the 3-form dark energy for early-time parameters under the influence of the late-time observation.}
\end{figure}

\begin{figure}
\centering
\subfigure[\label{fig:fit_CMBplus_late_params}Cumulatively larger CMB-based dataset]{\includegraphics[width = 0.48\textwidth]{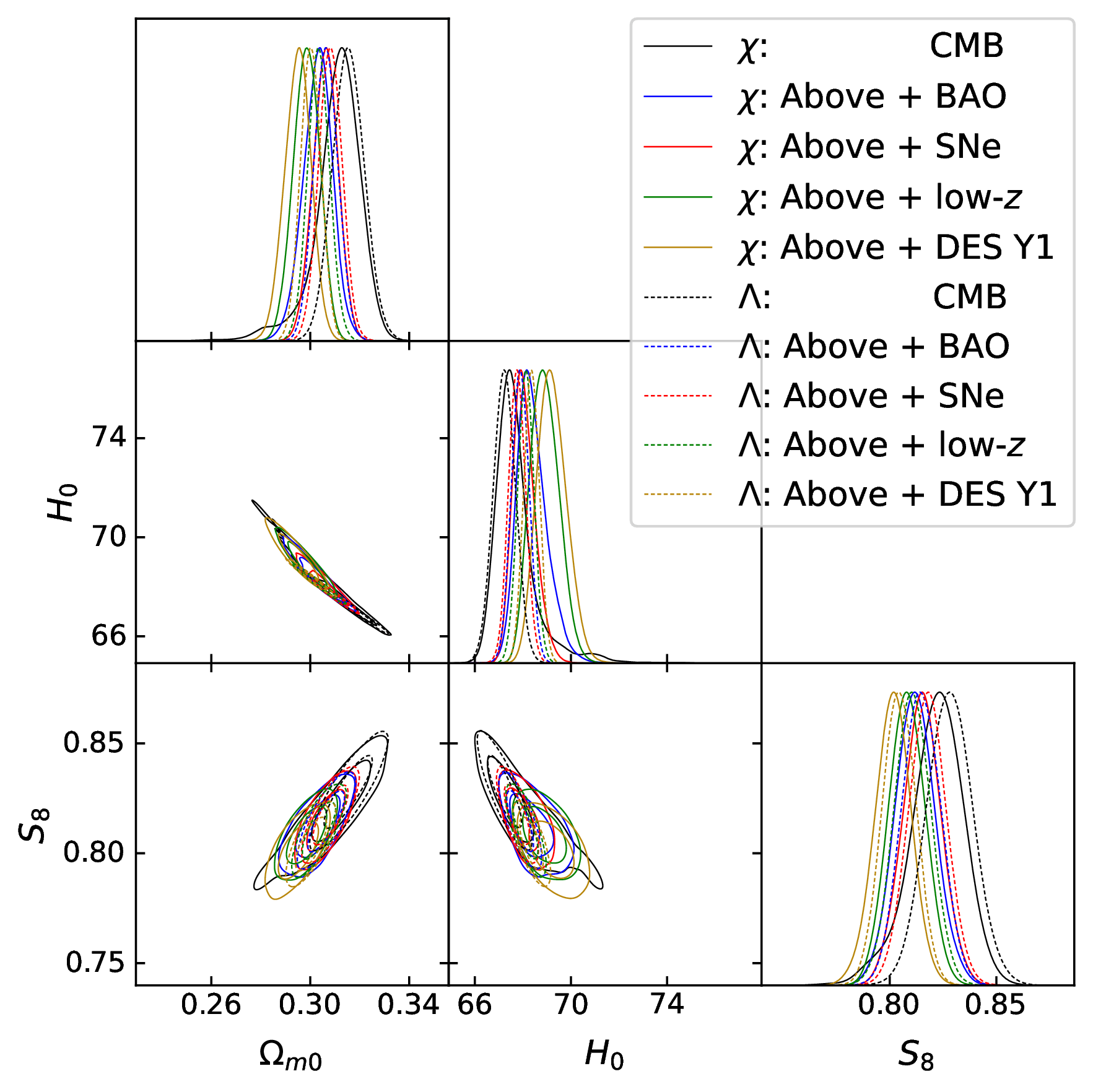}}~
\subfigure[\label{fig:fit_individual_late_params}Effect of 3-form inferred from individual dataset]{\includegraphics[width = 0.48\textwidth]{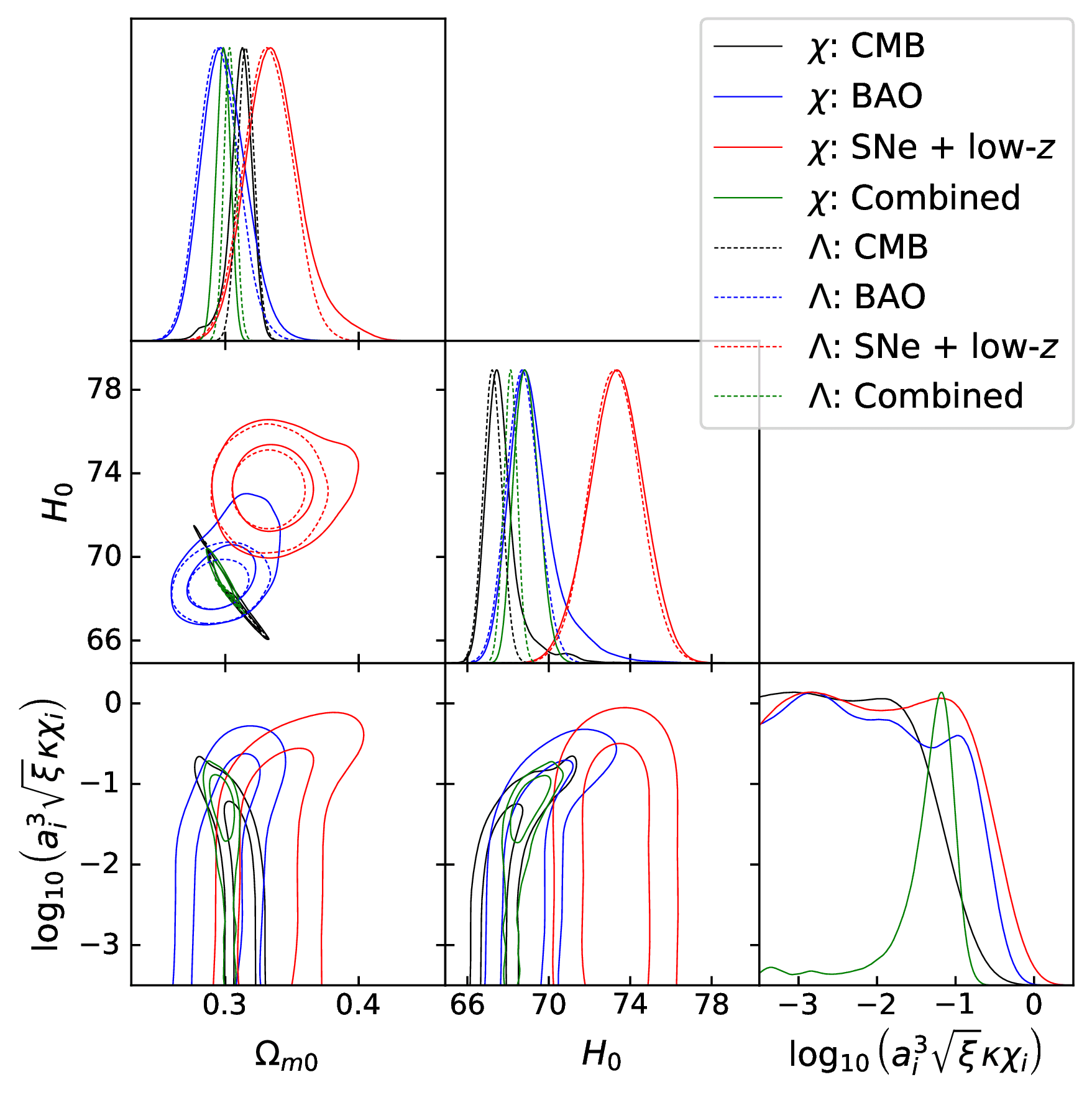}}
\caption{\label{fig:fit_late_params}
68\% and 95\% C.L. posterior distribution of the late-time quantities, with $\Lambda$CDM model ($\Lambda$) the dashed contours and the 3-form dark energy model ($\chi$) the solid contours. \emph{Left}: Inferred from CMB (black), CMB + BAO ({\color{RoyalBlue}blue}), CMB + BAO + SNe ({\color{BrickRed}red}), CMB + BAO + SNe + low-$z$ ({\color{ForestGreen}green}), and CMB + BAO + SNe + low-$z$ + DES Y1 ({\color{Dandelion}yellow}) data. \emph{Right}: Inferred from CMB (Black), BAO ({\color{RoyalBlue}blue}), SNe + low-$z$ ({\color{BrickRed}red}), and the combined ({\color{ForestGreen}green}, identical to the one on \emph{left}) data. The strong tension between astrophysical data of SNe + low-$z$ and cosmological data of CMB + BAO presented in $\Lambda$CDM model posterior of $H_0$ is moderately reduced in the 3-form dark energy model.
}
\end{figure}

Furthermore, as shown in Fig.~\ref{fig:fit_CMBplus_early_params}, the posterior distribution of the early-time cosmological parameters given different datasets are more clustered together for the 3-form model than $\Lambda$CDM model, signalling a reduction of the tension between early-time and late-time observation.

Meanwhile, as the posterior distribution of the Hubble parameter predicted by the 3-form model extends toward higher values relative to the prediction of $\Lambda$CDM model, the predicted value of $\Omega_{\rm m0}$ drops, as clearly demonstrated in Fig.~\ref{fig:fit_CMBplus_late_params}. This is common to all late time dark energy model as the matter density at the present time is strongly constrained by CMB observation. However, it does lead to an additional tension as high-redshift SNe provide direct constraint on $\Omega_{\rm m0}$.

\begin{table}
\centering
\scriptsize
\begin{tabular}{|c||c|c|c|}
\hline
                    & BAO                   & SNe + low-$z$         & DES Y1                \\
\hhline{|=#=|=|=|}
$10^3\Omega_{\rm c0}h^2$& $116.7    \pm8.3$ & $155      \pm11$      & $\color{blue}{\color{black}86}^{+23}_{\color{black}-10}$  \\
$10^5\theta_{\rm MC}$   & $1042 \pm10$      & $1090     \pm11$      & $1005^{+45}_{-34}$    \\
$\ln (10^{10} A_s)$ &                       &                       & $3.41^{+0.30}_{-0.26}$ \\
\hline
$H_0$               & $68.71    \pm0.80$    & $73.2     \pm1.3$     & $65.9     \pm7.1$     \\
$\Omega_{\rm m0}$   & $0.295    \pm0.015$   & $0.332    \pm0.018$   & $\color{blue}{\color{black}0.250}^{+0.040}_{\color{black}-0.032}$ \\
$\sigma_8$          &                       &                       & $0.845    \pm0.086$   \\
$S_8$               &                       &                       & $0.791    \pm0.029$   \\
\hline
DIC                 & $ 8.45    \pm0.05$    & $703.37   \pm0.03$    & $260.64   \pm0.75$    \\
WAIC                & $ 8.51    \pm0.15$    & $703.31   \pm0.03$    & $262.63   \pm0.10$    \\
$-\ln B$            & $ 8.86    \pm0.34$    & $703.49   \pm0.11$    & $262.35   \pm0.87$    \\
\hline
Tension against     & CMB                   & CMB + BAO             & CMB + BAO + SNe + low-$z$ \\
\hline
$-\ln R$            & $ 0.0     \pm1.4$     & $ 7.5     \pm1.6$     & $-1.6     \pm2.1$     \\
GoF                 & $ 1.94 \pm0.30\sigma$ & $ 4.30 \pm0.20\sigma$ & $ 2.54 \pm0.30\sigma$ \\
$S$                 & $ 1.68 \pm0.18\sigma$ & $ 4.06 \pm0.17\sigma$ & $ 1.60 \pm0.07\sigma$ \\
\hline
\end{tabular}
\caption{\label{table:tension_LambdaCDM} Mean and standard deviation of cosmological parameters, late-time observables, and statistical probes for $\Lambda$CDM model. From left to right are datasets of BAO, SNe + low-$z$, and DES Y1. Tension probes of $-\ln R$, GoF and $S$ are with respect to $\Lambda$CDM model inside table~\ref{table:result_LambdaCDM} according to ``Tension against'' row. For parameters not following Gaussian distribution we provide the median and 68\% lower and upper bounds (if valid) instead, with colour coding for how heavy the tail is ({\color{BrickRed}red} for short tail, black for Gaussian, {\color{blue}blue} for exponential, and {\color{Cyan}cyan} for long tail.)}
\end{table}

\begin{table}
\centering
\scriptsize
\begin{tabular}{|c||c|c|c|}
\hline
                            & BAO           & SNe + low-$z$ & DES Y1        \\
\hhline{|=#=|=|=|}
$10^3 \Omega_{\rm c0} h^2$  & $121  \pm12$          & $159  \pm14$          & $\color{blue}{\color{black}86}^{+14}_{\color{black}-15}$  \\
$10^5 \theta_{\rm MC}$  & $\color{blue}{\color{black}1044}^{+11}_{\color{black}-11}$& $1089 \pm12$  & $1002 \pm43$  \\
$\ln ( 10^{10} A_s )$       &                       &                       & $\color{white}{\color{black}3.52}^{+?}_{\color{black}-0.39}$  \\
\hline
$\log_{10} \xi$         & $\color{cyan}-3.6$    & $\color{cyan}-3.4$    & $\color{cyan}-3.6$    \\
$({\rm KE}_i + V_\ast) / \rho_{\rm DE,0}$   & $1.0002\pm{\color{cyan}0.0027}$   & $1.0003\pm{\color{cyan}0.0041}$   & $1.0001\pm{\color{cyan}0.0025}$   \\
$\log_{10} (a_i^3 \sqrt\xi \kappa \chi_i)$  & $-2.1^{+0.7}_{\color{white}-?}$   & $\color{cyan}-1.9$    & $-2.0^{+0.5}_{\color{white}-?}$   \\
$(\dot\chi_i + 3 H_i \chi_i) \rho_{\rm DE,0}^{-1/2}$  & $ 0.03\pm0.61$& $0.03\pm0.61$ & $0.10\pm0.63$ \\
\hline
$H_0$   & $\color{blue}{\color{black}68.9}^{+1.0}_{\color{black}-1.1}$  & $73.3 \pm1.3$ & $66.4 \pm7.2$ \\
$\Omega_{\rm m0}$   & $\color{blue}{\color{black}0.298}^{+0.017}_{\color{black}-0.017}$ & $\color{blue}{\color{black}0.335}^{+0.019}_{\color{black}-0.022}$ & $\color{blue}{\color{black}0.264}^{+0.041}_{\color{black}-0.044}$ \\
$\sigma_8$                  &                       &                       & $0.845    \pm0.086$   \\
$S_8$                       &                       &                       & $0.791    \pm0.029$   \\
\hline
DIC                         & $ 8.54    \pm0.05$    & $703.70   \pm0.06$    & $261.51   \pm0.77$    \\
WAIC                        & $ 8.25    \pm0.04$    & $703.45   \pm0.06$    & $263.5    \pm1.1$     \\
$-\ln B$                    & $ 8.57    \pm0.19$    & $703.53   \pm0.33$    & $263.00   \pm0.98$    \\
\hline
$\Delta$DIC                 & $ 0.09    \pm0.08$    & $  0.33   \pm0.07$    & $  0.9    \pm1.1$     \\
$\Delta$WAIC                & $-0.26    \pm0.18$    & $  0.13   \pm0.07$    & $  0.9    \pm1.1$     \\
$-\Delta\ln B$              & $-0.29    \pm0.40$    & $  0.04   \pm0.36$    & $  0.6    \pm1.4$     \\
\hline
Tension against             & CMB                   & CMB + BAO             & CMB + BAO + SNe + low-$z$ \\
\hline
$R$                         & $ -0.2    \pm1.8$     & $ 4.3     \pm1.3$     & $-3.4     \pm1.1$     \\
GoF                         & $ 1.95 \pm0.50\sigma$ & $ 4.04 \pm0.28\sigma$ & $ 2.31 \pm0.51\sigma$ \\
$S$                         & $ 1.77 \pm0.49\sigma$ & $ 3.63 \pm0.34\sigma$ & $ 1.42 \pm0.22\sigma$ \\
\hline
$\Delta R$                  & $ -0.2    \pm2.3$     & $-3.2     \pm2.1$     & $-1.8     \pm2.4$     \\
$\Delta{\rm GoF}$           & $ 0.01 \pm0.58\sigma$ & $-0.26 \pm0.34\sigma$ & $-0.23 \pm0.59\sigma$ \\
$\Delta S$                  & $ 0.09 \pm0.52\sigma$ & $-0.43 \pm0.38\sigma$ & $-0.18 \pm0.23\sigma$ \\
\hline
\end{tabular}
\caption{\label{table:tension_scalar} Mean and standard deviation of cosmological parameters, late-time observables, and statistical probes for the 3-form dark energy model in section~\ref{sec:model}. From left to right are datasets of BAO, SNe + low-$z$, and DES Y1. $\Delta{\rm ICs}$ and delta of tension probes are with respect to $\Lambda$CDM model presented in table~\ref{table:tension_LambdaCDM}. Tension probes of $-\ln R$, GoF and $S$ are with respect to axion-like dark energy model inside table~\ref{table:result_scalar} according to ``Tension against'' row. For parameters not following Gaussian distribution we provide the median and 68\% lower and upper bounds (if valid) instead, with colour coding for how heavy the tail is ({\color{BrickRed}red} for short tail, black for Gaussian, {\color{blue}blue} for exponential, and {\color{Cyan}cyan} for long tail.) If the distribution is clearly single-sided we report the modal and the single-sided 68\% bound instead.}
\end{table}

To quantitatively estimate this trade-off between the Hubble tension and the ``$\Omega_{\rm m0}$ tension,'' we utilise the statistical analysis described at the beginning of section~\ref{sec:fit}. We perform the MCMC analyses of BAO, SNe+low-$z$ and DES Y1 datasets alone without CMB dataset. Mean and standard deviation of the cosmological parameters, late-time observables, and statistical probes are shown in table~\ref{table:tension_LambdaCDM} for the $\Lambda$CDM model and in table~\ref{table:tension_scalar} for the 3-form dark energy model. Notice that the lack of constraint on $\sigma_8$ and $S_8$ for BAO and SNe + low-$z$ datasets is expected, as we impose no a priori constraint on the primordial perturbation. Similarly, $n_s$ and $\tau_{\rm reio}$ are omitted due to the lack of constraint by the late-time physics. $\Omega_{\rm b0} h^2$ is omitted as well since it would only reflect the BBN prior. In addition, the DES Y1 posterior for the 3-form dark energy model is a rough estimate, as we cannot reach the stopping criteria. 

We then carry out the statistical probes of Bayesian ratio, goodness of fit and suspiciousness, and find that the 3-form dark energy model overall shows clear statistical advantage over $\Lambda$CDM model in terms of data tension between CMB + BAO dataset and SNe + low-$z$ dataset, and no apparent (dis-)advantage when confronting CMB against BAO, or CMB + BAO + SNe + low-$z$ against DES weak lensing dataset. This is consistent with the theoretical prediction that the effect of the 3-form is rather subtle except for the background Hubble expansion. We explicitly demonstrate this in Fig.~\ref{fig:fit_individual_late_params}, where by tuning the initial field strength we can reduce the tension between CMB and SNe + low-$z$ dataset. To compensate the phantom-ness of the 3-form, the dark matter fraction has to be higher than anticipated to explain the observed SNe luminosity distance, and lower to maintain the CMB constraint on matter density. Together, we find a window of $a_i^3 \sqrt\xi \kappa \chi_i \sim 10^{-1}$.

\begin{figure}
\centering
\subfigure[\label{fig:fit_scatter_CMB}CMB]{\includegraphics[width = 0.495 \textwidth]{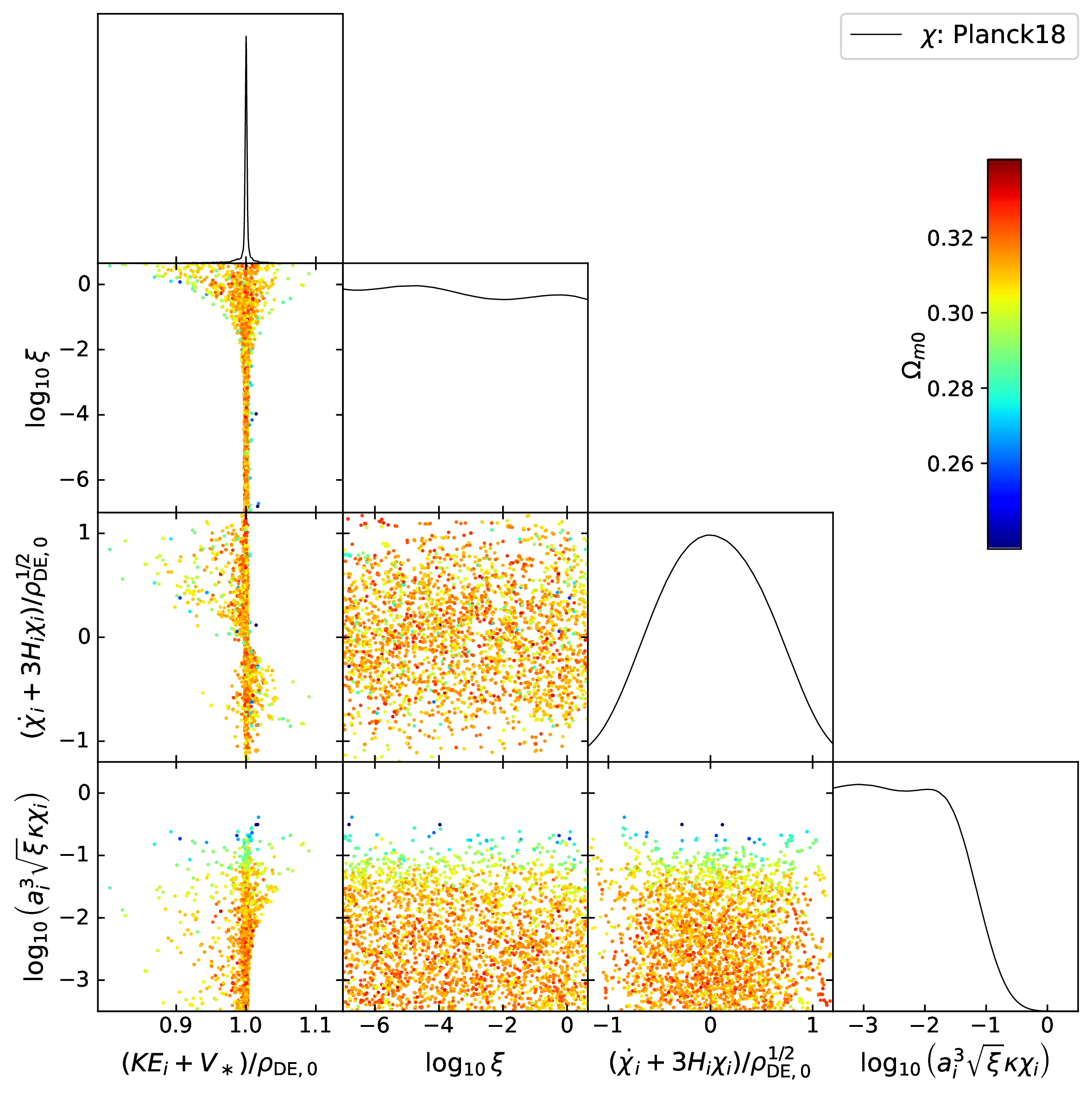}}
\subfigure[\label{fig:fit_scatter_SN_LowZ}SNe + low-$z$]{\includegraphics[width = 0.495 \textwidth]{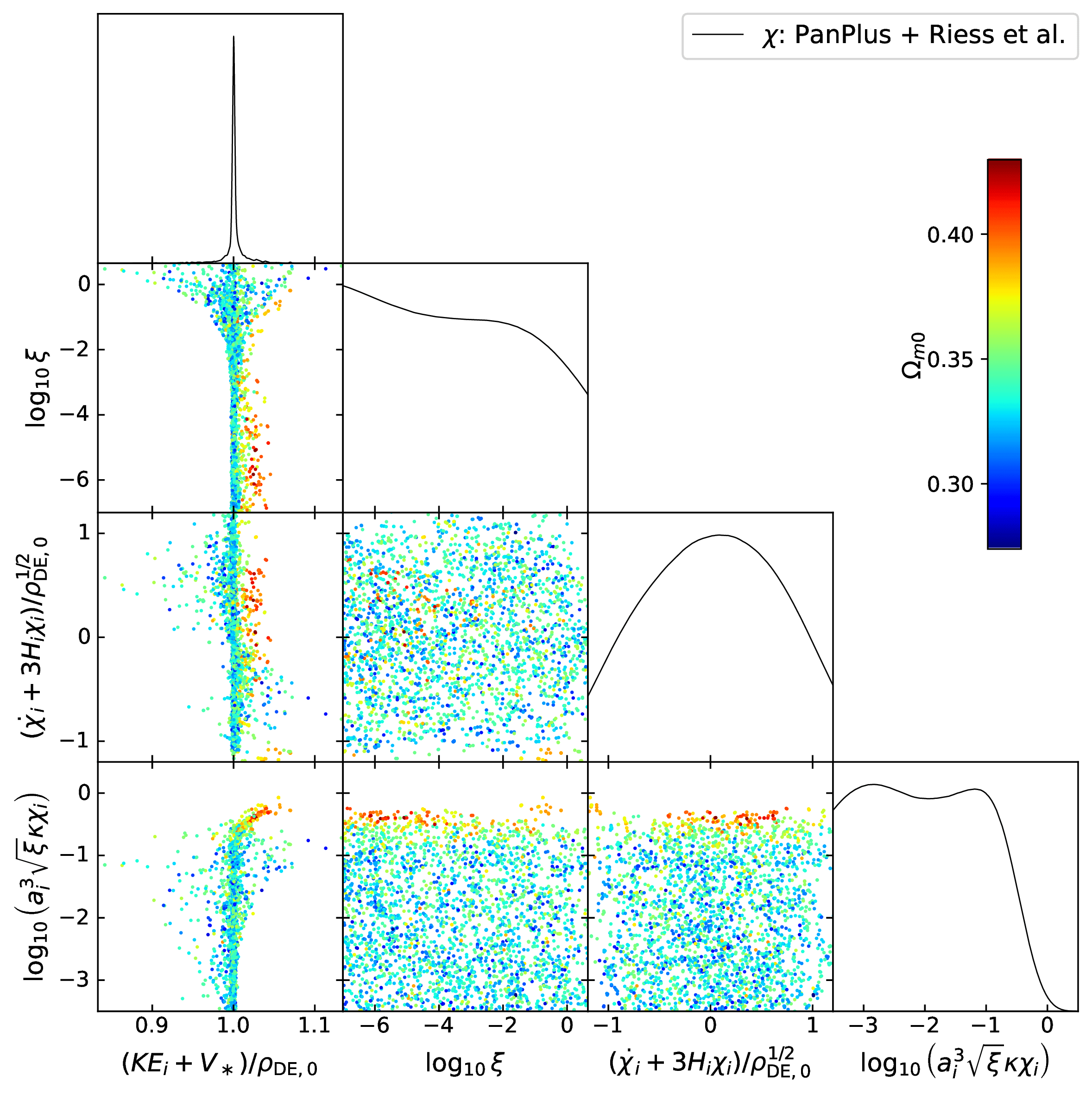}}
\caption{\label{fig:scatter}
Scatter plot of the extended parameters for the 3-form dark energy model, colour-coded by $\Omega_{\rm m0}$, with CMB on the \emph{top} and SNe + low-$z$ on the \emph{bottom}.
}
\end{figure}

To better understand how the 3-form dark energy model affects the cosmological fit, the extended parameters of the model are shown in Figs.~\ref{fig:scatter} for CMB and SNe + low-$z$ dataset as scatter plots over the current matter density parameter $\Omega_{\rm m0}$. We choose to omit $H_0$ as it is either fully constrained (SNe + low-$z$) or strongly correlated with $\Omega_{\rm m0}$ (CMB), thus providing no additional information.

One can clearly infer from the figure that only the initial field strength influences the cosmic evolution. This is consistent with the assumption that the 3-form is driven by the Hubble flow for the majority of the cosmic evolution, and its effect is well characterised by when it climbs up the potential. Assuming that the 3-form is driven entirely by the Hubble flow ($a \chi^3 \sim \text{constant}$), the climbing of the potential ends at $\sqrt{\xi/3} \kappa \chi \sim a_i^3\sqrt{\xi/3} \kappa \chi_i a^{-3} \sim 0.05 a^{-3} \sim 1$, i.e., at $z\sim 1.7$, where the effect on the distance to the last scattering surface is maximised while the effect on high-$z$ SNe is minimised as SNe in Pantheon+ catalogue are mostly at $z<1.7$.\footnote{The choice  of \COMREV{1-$\sigma$ deviation from the centre}\REV{the inflection point} of the Gaussian potential as the end of climbing is arbitrary. However, it does correspond to the transition from $c_s^2 < 0$ to $c_s^2 > 0$. Therefore, the perturbation stabilises during the dark-matter-dark-energy transition\COMREV{, avoiding the materialisation of the gradient instability}.\REV{~Before that, one should consider the 3-form field as merely an effective description.}} This is analogous to the phenomenological sign-switching $\Lambda_s$CDM model \cite{Akarsu:2021fol, Akarsu:2022typ, Akarsu:2023mfb, Bouhmadi-Lopez:2025ggl}, where a sudden change to the cosmological constant at $z \sim 2$ splits the universe into a standard late-time epoch with $\Omega_\Lambda \sim 0.7$ and a new high-$z$ epoch with a much lower $\Omega_\Lambda$. A natural extension to the \REV{phantom }3-form model \COMREV{with a Gaussian potential which is}\REV{currently} under investigation thus would be the inclusion of a negative cosmological constant, as it may be regarded as a theoretically well-motivated realisation of sign-switching $\Lambda_s$CDM\REV{~if one can mend the gradient instability issue}.

\subsection{Cosmic evolution, CMB power spectrum, matter power spectrum and $f\sigma_8$}
\label{sec:evolution}

\begin{figure}
\centering
\subfigure[CMB + BAO: $f \sigma_8$\label{fig:fsigma8_CMB_BAO}]{\includegraphics[width = 0.49 \textwidth, trim = 0 10 0 2015, clip]{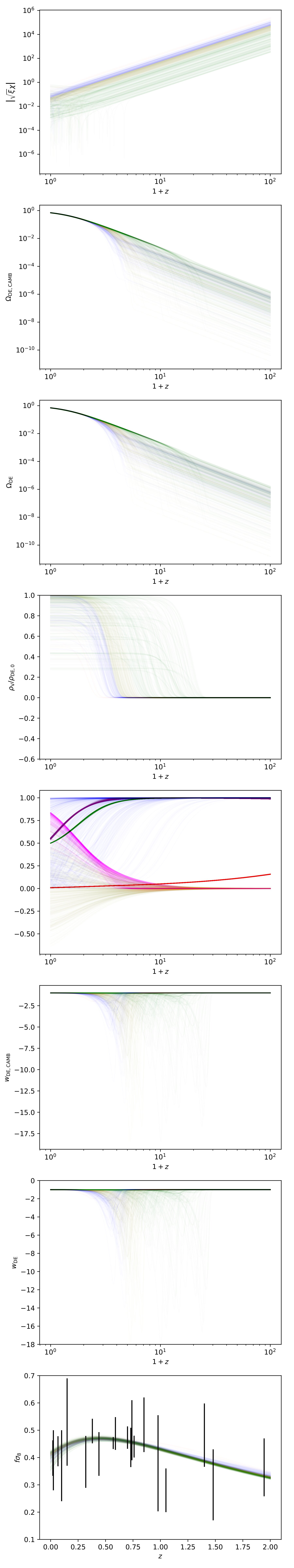}}~
\subfigure[CMB + BAO + SNe + low-$z$: $f \sigma_8$\label{fig:fsigma8_CMB_BAO_SN_LowZ}]{\includegraphics[width = 0.49 \textwidth, trim = 
0 10 0 2015, clip]{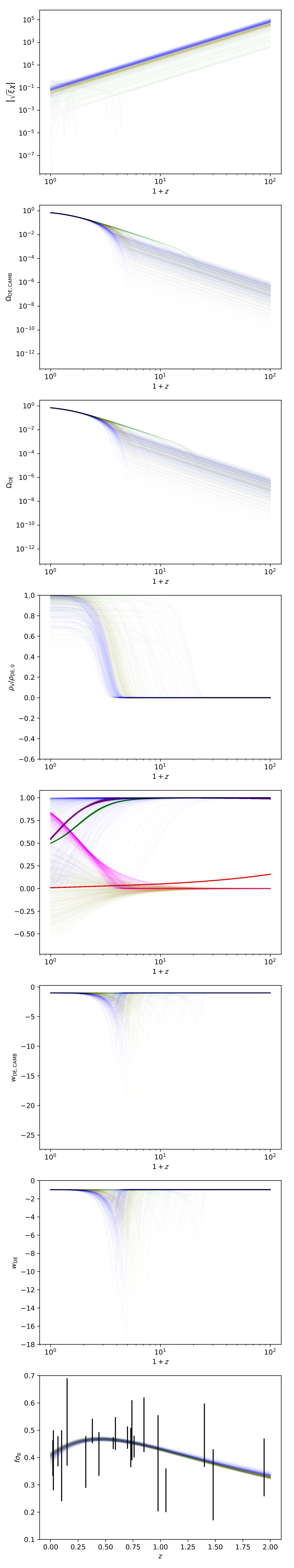}}
\\
\subfigure[CMB + BAO: $w_{\rm DE}$\label{fig:w_CMB_BAO}]{\includegraphics[width = 0.49 \textwidth, trim = 0 295 0pt 1730, clip]{test_ThreeForm_Gaussian_Lambda_0_CMB_Lens_DESI.jpg}}~
\subfigure[CMB + BAO + SNe + low-$z$: $w_{\rm DE}$\label{fig:w_CMB_BAO_SN_LowZ}]{\includegraphics[width = 0.49 \textwidth, trim = 
0 295 0 1730, clip]{test_ThreeForm_Gaussian_Lambda_0_CMB_Lens_DESI_PanPlus_Riess.jpg}}
\\
\subfigure[CMB + BAO: variables\label{fig:dyn_var_CMB_BAO}]{\includegraphics[width = 0.49 \textwidth, trim = 0 870 0 1160, clip]{test_ThreeForm_Gaussian_Lambda_0_CMB_Lens_DESI.jpg}}~
\subfigure[CMB + BAO + SNe + low-$z$: variables\label{fig:dyn_var_CMB_BAO_SN_LowZ}]{\includegraphics[width = 0.49 \textwidth,trim = 
0 870 0 1160, clip]{test_ThreeForm_Gaussian_Lambda_0_CMB_Lens_DESI_PanPlus_Riess.jpg}}
\caption{\label{fig:cosmic_evolution}
$f\sigma_8$ (\emph{top}), dark energy EoS $w_{\rm DE}$ (\emph{centre}) and dynamical variables (\emph{bottom}) as functions of $z$ for 600 samples drawn from individual fits, against CMB + BAO (\emph{left}) and CMB + BAO + SNe + low-$z$ (\emph{right}) dataset. In \emph{bottom} figures, colour represents variables of section~\ref{sec: dyn_var}: {\color{RoyalBlue}blue}: $v$, {\color{Purple}purple}: $s$, {\color{ForestGreen}green}: $\mathfrak h$, {\color{BrickRed}red}: $r$, {\color{Dandelion}yellow}: $y$, {\color{magenta}magemta}: $\kappa\sqrt{V/3}/H$. In \emph{top} and \emph{centre} figures, the colour indicates the initial field strength $\log_{10}(a_i^3 \sqrt\xi \kappa \chi_i)$: ({\color{BrickRed}red}: $>-3/4$, {\color{RoyalBlue}blue}: $-3/4 \sim -5/4$, {\color{Dandelion}yellow}: $-5/4 \sim -7/4$, {\color{ForestGreen}green}: $<-7/4$). In \emph{top} figures, the $f\sigma_8$ data are taken from table~2 of \cite{Avila:2022xad}.
}
\end{figure}

To better understand the effect of the 3-form on the cosmological observable, we show the evolution of the dynamical variables introduced in section~\ref{sec: dyn_var}, the dark energy EoS $w_{\rm DE}$, and $f\sigma_8$ from the posterior of the 3-form model for CMB + BAO and CMB + BAO + SNe + low-$z$ dataset in Figs.~\ref{fig:cosmic_evolution}.

First, the cross-over of importance between the kinematic term and the Hubble flow occurs at $3 H \chi \sim 3 \kappa^{-1} H \xi^{-1/2} a^{-3} \left( a_i^3 \sqrt \xi \kappa \chi_i \right) \sim \sqrt{2{\rm KE}_i} \sim \sqrt{3} \kappa^{-1} H_0 \sqrt{\Omega_{\rm DE,0}} \sqrt{2{\rm KE}_i/\rho_{\rm DE,0}}$, where we have assumed the dominance of the Hubble flow $a^3\chi \sim a_i^3 \chi_i$ and ${\rm KE} \sim {\rm KE}_i$. Thus it is clear that $a_i^3 \sqrt \xi \kappa \chi_i$, $\xi$ and ${\rm KE}_i$ together control the cross-over, and in turn control when we are repulsed by the de Sitter fixed point $C$. The prior space we consider spans both scenarios of moving toward or away from $C$ at the present time, as shown by the {\color{RoyalBlue}blue} curves ($v / (1 - v^2) \equiv \chi /\chi_c$) in Figs.~\ref{fig:dyn_var_CMB_BAO} and \ref{fig:dyn_var_CMB_BAO_SN_LowZ}. But since the eventual LSBR attractor takes an infinite amount of time to approach, observationally it is difficult to distinguish between the two. Thus, the preference is mostly determined by the prior volume effect.

Still, one can conclude that the preferably larger value of $a_i^3 \sqrt \xi \kappa \chi_i$ from the inclusion of SNe + low-$z$ dataset leads to a preference of moving toward the fixed point $C$ at the present time, as shown in Figs.~\ref{fig:dyn_var_CMB_BAO} and \ref{fig:dyn_var_CMB_BAO_SN_LowZ} where $v$ ({\color{RoyalBlue}blue}) tends closer to $1$ and $y$ closer to $0$ when SNe + low-$z$ is included (\emph{right}). However, due to the choice of the flat prior in $(\dot\chi_i + 3 H_i\chi_i) / \rho_{{\rm crit,}h=0.7}^{-1/2} \sim y_0$, it is natural that we span rather evenly between $E_+^\pm$ and $C$.

Notice that this difference in which de Sitter fixed point we are tending toward or moving away does not affect when we are climbing the potential, as evidently shown in Figs.~\ref{fig:w_CMB_BAO} and \ref{fig:w_CMB_BAO_SN_LowZ}, where we colour the curves according to the initial field strength $\log_{10}(a_i^3 \sqrt\xi \kappa \chi_i)$, {\color{BrickRed}red} if larger than $-3/4$, {\color{RoyalBlue}blue} if between $-3/4$ and $-5/4$, {\color{Dandelion}yellow} if between $-5/4$ and $-7/4$, and {\color{ForestGreen}green} if smaller than $-7/4$. As discussed previously, the inclusion of the SNe + low-$z$ dataset pinpoints the initial field strength, which in turn dictates when we should reach the top of the potential, as shown by the preference of $w_{\rm DE}$ deviating from $-1$ at $z \sim 2$. Yet, there is no visible effect on the evolution of the dynamical variables, other than the matter density parameter $s^2$ dropping faster during the climb. It does affect the matter growth history though, as shown in Figs.~\ref{fig:fsigma8_CMB_BAO} and \ref{fig:fsigma8_CMB_BAO_SN_LowZ}. At $z < 1.7$, $f\sigma_8$ remains identical to the prediction of $\Lambda$CDM model, but there is a preference for a slight enhancement at a higher redshift for {\color{RoyalBlue}blue} curves than $\Lambda$CDM predictions (well characterised by {\color{ForestGreen}green} curves), affecting the matter power spectrum, as will be shown below.

\begin{figure}
\centering
\subfigure[CMB high-$\ell$ TT power spectrum\label{fig:power_spectrum_high_l}]{
\begin{tikzpicture}[      
        every node/.style={anchor=south west,inner sep=0pt},
        x=\textwidth, y=\textwidth,
      ]   
     \node (fig2) at (0,0)
       {\includegraphics[width = 0.475\textwidth]{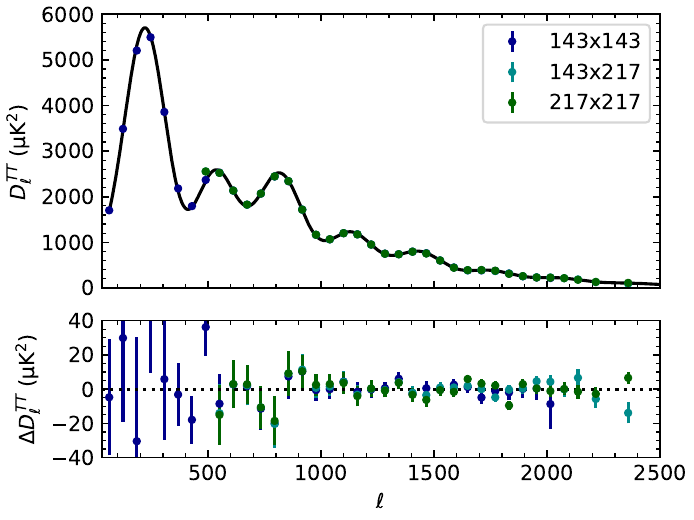}};  
     \node (fig1) at (0.009,0.0285)
       {\includegraphics[width = 0.462\textwidth, trim = 0cm 0cm 0.5cm 0cm, clip]{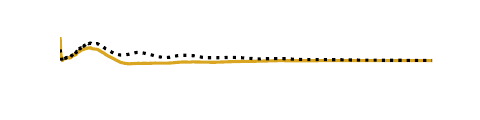}};
     \node (fig3) at (0.009,0.0285)
       {\includegraphics[width = 0.462\textwidth, trim = 0cm 0cm 0.5cm 0cm, clip]{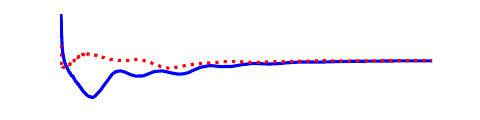}};
\end{tikzpicture}
}
\subfigure[CMB low-$\ell$ TT power spectrum\label{fig:power_spectrum_lowl}]{
\begin{tikzpicture}[      
        every node/.style={anchor= west,inner sep=0pt},
        x=\textwidth, y=\textwidth,
      ]   
     \node (fig2) at (0,0)
       {\includegraphics[width = 0.475\textwidth, trim = 0 0.2cm 0 0, clip]{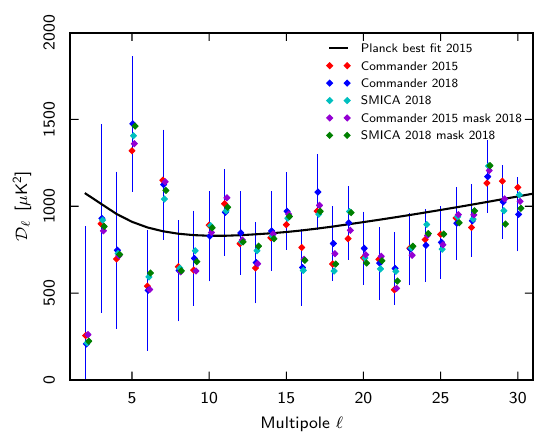}};
     \node (fig1) at (-0.0005,0.0115)
       {\includegraphics[width = 0.475\textwidth, trim = 0.1cm 0.1cm 1.1cm 0.1cm, clip]{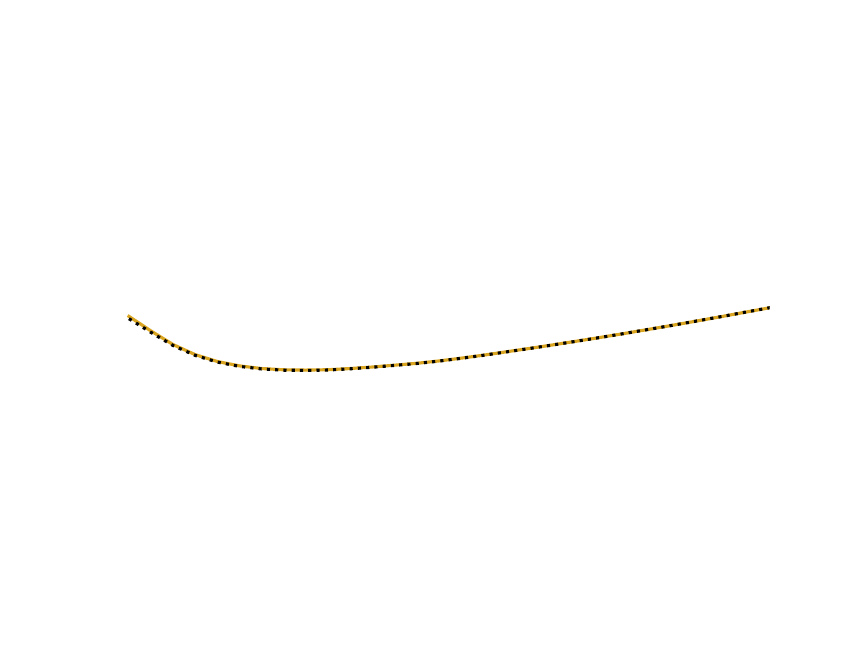}};
     \node (fig3) at (-0.0005,0.0115)
       {\includegraphics[width = 0.475\textwidth, trim = 0.1cm 0.1cm 1.1cm 0.1cm, clip]{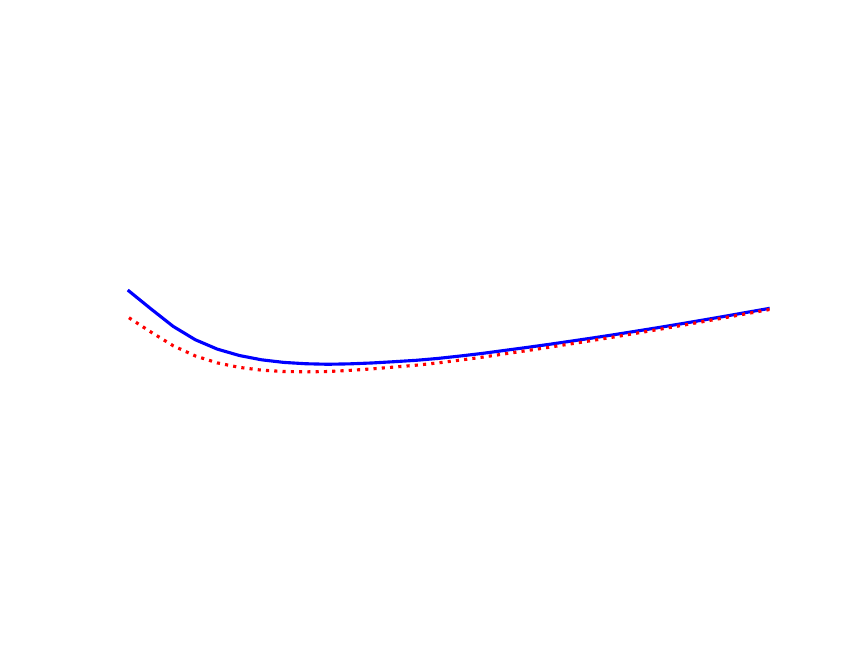}};
\end{tikzpicture}
}
\caption{
\emph{Top left}: Best fit TT power spectrum of $\Lambda$CDM model against Planck CamSpec PR4 high-$\ell$ TTTEEE dataset and the PR4 high-$\ell$ TT data points in Fig.~6 of \cite{Rosenberg:2022sdy}. \emph{Bottom left}: Residue of the best fit TT power spectrum and the PR4 high-$\ell$ TT data points against best fit $\Lambda$CDM model of Planck CamSpec PR4 high-$\ell$ TTTEEE dataset. \emph{Right}: Best fit CMB TT low-$\ell$ power spectra and the data of various maps and masks from Planck18 data release in Fig.~2 of \cite{Planck:2019nip}. Black solid line is the best fit $\Lambda$CDM model prediction from Planck15.
Black \emph{dotted} line and {\color{BrickRed}red} \emph{dotted} line denote $\Lambda$CDM model fitted against CMB and CMB + BAO datasets respectively, and {\color{Dandelion}yellow} \emph{solid} line and {\color{RoyalBlue}blue} \emph{solid} line denotes 3-form dark energy model fitted against CMB and CMB + BAO datasets respectively. There are no significant difference between each case, except for the 3-form model against CMB + BAO dataset ({\color{RoyalBlue}blue}) at high-$\ell$. The difference between the CMB-only best fit $\Lambda$CDM model in our analysis and the Planck CamSpec PR4 high-$\ell$ TTTEEE best fit model in \cite{Rosenberg:2022sdy} stems from the difference in data inclusion, as we also include low-$\ell$ TTEE dataset \cite{Planck:2019nip} and PR4 CMB lensing dataset \cite{Carron:2022eyg, Carron:2022eum}.
}
\end{figure}

\begin{figure}
\centering
\begin{tikzpicture}[      
        every node/.style={anchor=south west,inner sep=0pt},
        x=\textwidth, y=\textwidth,
      ]   
     \node (fig2) at (0,0)
       {\includegraphics[width = 0.6\textwidth]{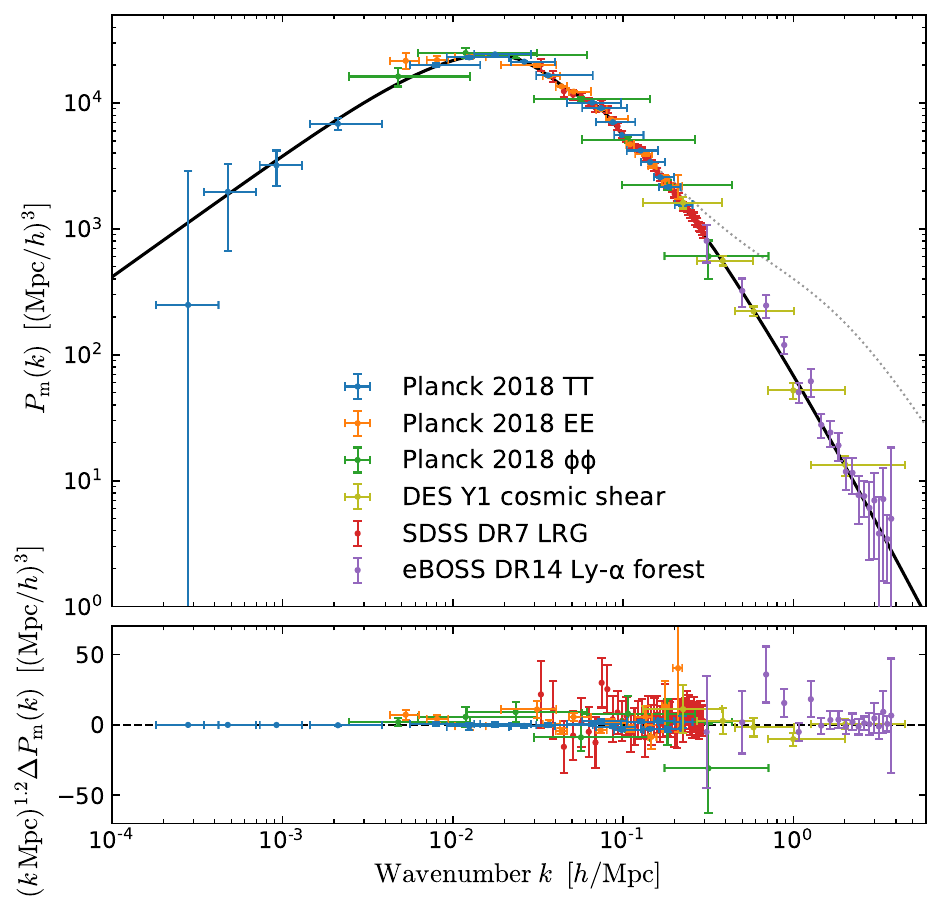}};
     \node (fig1) at (0.002,0.042)
       {\includegraphics[width = 0.6\textwidth, trim = 0.25cm 0 1cm 0,clip]{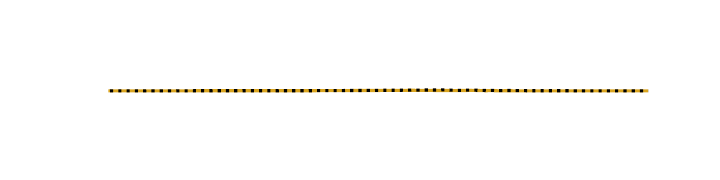}};
     \node (fig3) at (0.002,0.042)
       {\includegraphics[width = 0.6\textwidth, trim = 0.25cm 0 1cm 0,clip]{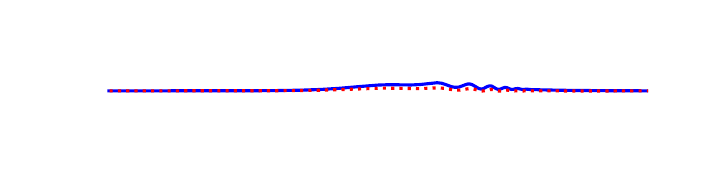}};
\end{tikzpicture}
\caption{\label{fig:matter_power_spectrum}
The best fit matter power spectrum (\emph{top}) and the residue (\emph{bottom}) against best fit $\Lambda$CDM model matter power spectrum of CMB dataset. \emph{Top}: \emph{Solid} lines represents the linear power spectrum and the \emph{dotted} lines represents the nonlinear power spectrum. Matter power spectrum data taken from Fig.~1 of \cite{Chabanier:2019eai}. \emph{Bottom}: Black \emph{dashed} line and {\color{BrickRed}red} \emph{dotted} line denote $\Lambda$CDM model fitted against CMB and CMB + BAO datasets respectively, and {\color{Dandelion}yellow} \emph{solid} line and {\color{RoyalBlue}blue} \emph{solid} line denotes the 3-form dark energy model fitted against CMB and CMB + BAO datasets respectively. While two models predict nearly identical matter power spectrum when considering only CMB dataset, there is a slight enhancement in the 3-form model once BAO data is included.
}
\end{figure}

\begin{figure}
\centering
\subfigure[$\Lambda$CDM]{\includegraphics[width=0.48\linewidth]{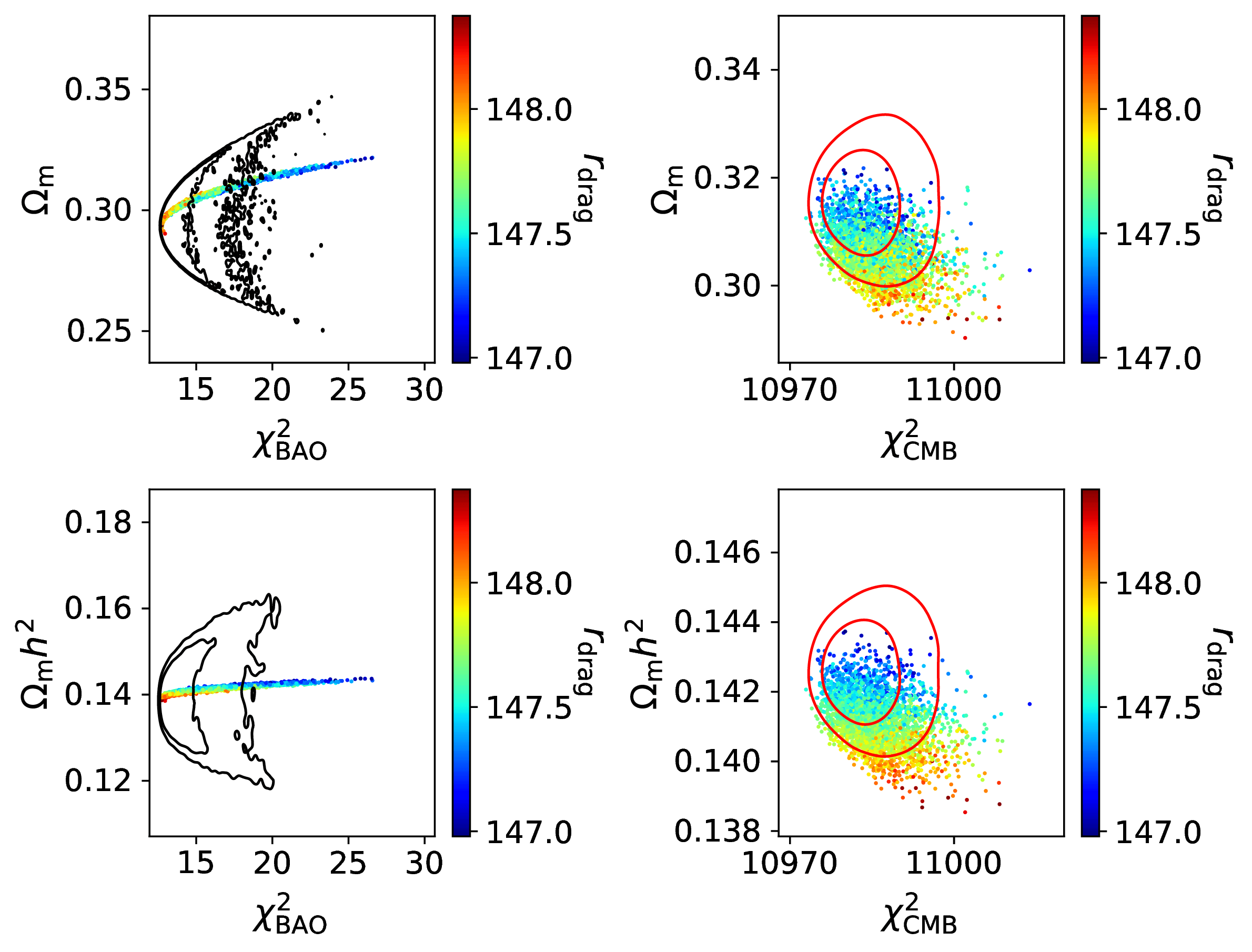}}
\subfigure[3-form]{\includegraphics[width=0.48\linewidth]{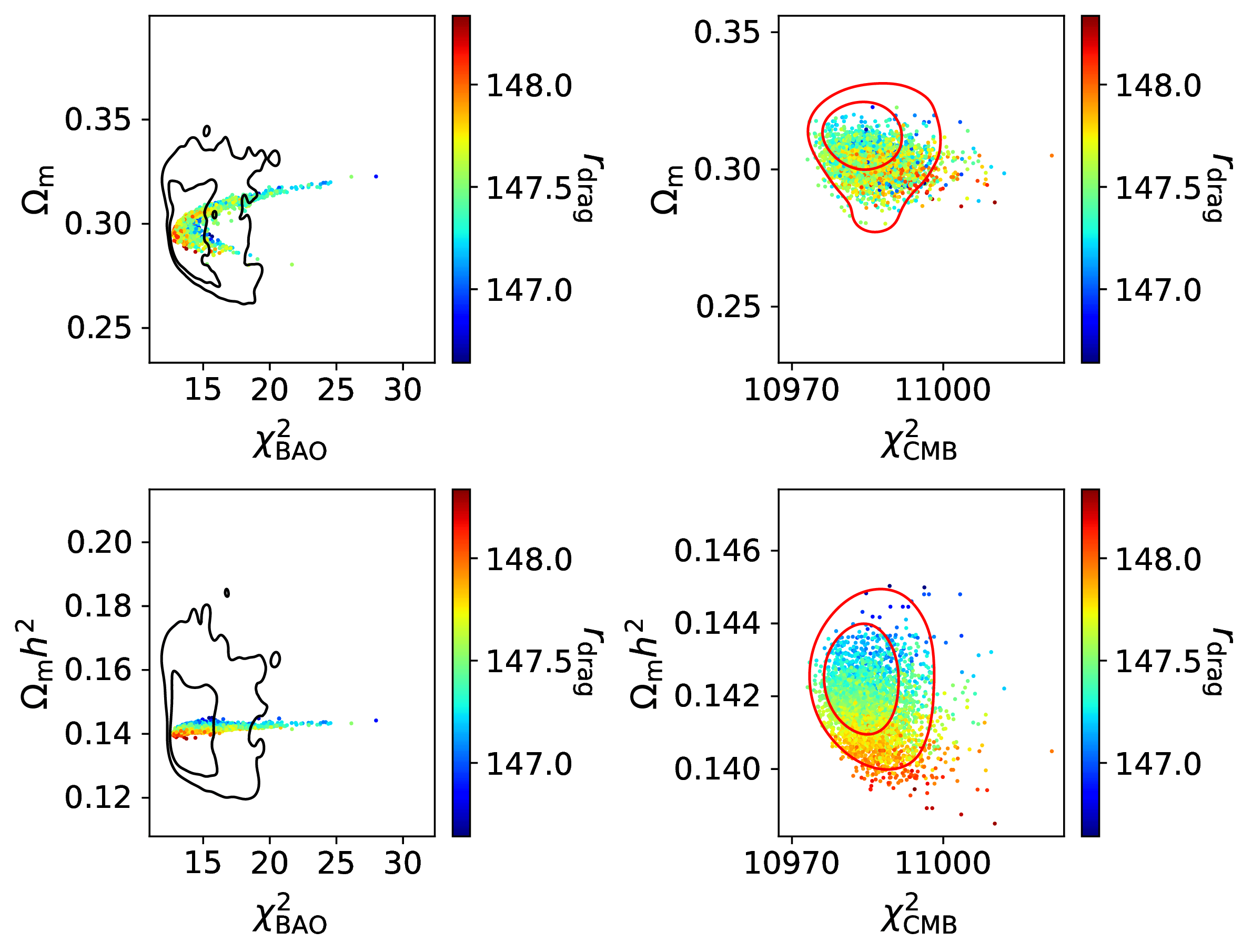}}
\caption{Scattering plots for $\Lambda$CDM (\emph{a}) and 3-form dark energy model (\emph{b}) fitted against CMB + BAO dataset, colour coded by the sound horizon at the drag scale $r_d$. For each sub-figure we present alongside the scattering plot, the 2-$d$ contours of the fit against BAO dataset on the \emph{left} in black, and against CMB dataset on the \emph{right} in {\color{BrickRed}red}. The tension in $r_d$ is clearly visible in both models, even when the $\Omega_{\rm m0}$, and accordingly the $H_0$ tension, is reduced in 3-form model given the relaxed late-time expansion history. This tension reduction leads to a slightly modified CMB power spectrum as shown in Fig.~\ref{fig:power_spectrum_high_l}.}
\label{fig:fit_rdrag}
\end{figure}

Finally, we isolate the maximum-a-posteriori prediction of the 3-form dark energy model and the $\Lambda$CDM model by the datasets of CMB and CMB + BAO and compare them in detail against the prediction of best fit $\Lambda$CDM model from Planck18. As shown in Figs.~\ref{fig:power_spectrum_high_l}, \ref{fig:power_spectrum_lowl} and Fig.~\ref{fig:matter_power_spectrum}, there are no clear difference between models from the perspective of CMB power spectrum and matter power spectrum when considering CMB dataset only. However, the prediction from the 3-form model with CMB + BAO dataset deviates from the rest, showing the compromise made when combining the distance measurement of BAO to CMB power spectrum. As suggested in Figs.~\ref{fig:fit_late_params}, this compromise comes from the relaxation of the expansion history at the late time, which was completely fixed by CMB data in $\Lambda$CDM model (as shown by $-\ln R \approx 0$ for BAO against CMB in Table.~\ref{table:tension_LambdaCDM}).\footnote{The statistic probes fail to identify this tension as it is rather weak. Furthermore, the 3-form model barely alleviates the main source of tension, i.e. the sound horizon at the drag scale $r_d$ \cite{DESI:2025fii}\REV{, as shown in Figs.~\ref{fig:fit_rdrag}}. Despite a relaxed expansion history, this $\Omega_{\rm m0}$-$H_0r_d$ tension between CMB, BAO and Riess $H_0$ measurement remains, showing up as an enhancement to matter spectrum power in unit of $({\rm Mpc}/h)^3$.} The figures for CMB + BAO + SNe + low-$z$ dataset are excluded as the result resembles those from the CMB + BAO dataset. This is expected given the consistency between the combined dataset and BAO data alone.

\section{Conclusions}
\label{sec:conclusion}

In this work, we first revisited 3-form dark energy models endowed with a potential and examined their key dynamical properties. We showed that such models can account for late-time cosmic acceleration through either quintessence-like or phantom-like behaviour, focusing in particular on the phantom-like regime arising from a 3-form field with a Gaussian potential, as we were interested in scenarios capable of alleviating the $H_0$ tension; something we indeed verified observationally. To gain a proper understanding of the model, we carried out a careful and comprehensive dynamical system analysis that incorporated radiation and dark matter alongside the 3-form dark energy sector equipped with a Gaussian potential. We found that the model exhibits an initial repulsive radiation phase, followed by a saddle matter-dominated phase, and subsequently a dark energy–dominated phase which, in the infinite future, evolves towards a Little Sibling of the Big Rip (LSBR), a considerably milder version of the classical Big Rip singularity. While in standard quintessence models both the radiation- and matter-dominated eras typically appear as saddle points, in the 3-form case with a Gaussian potential we recover the usual $\Lambda$CDM-like structure with a repulsive radiation point followed by a saddle matter point. We next address the observational viability of this model, and, to the best of our knowledge, this is the first time that 3-form fields have been observationally constrained as a dark energy model.

We conduct a thorough and careful MCMC analysis on a comprehensive observational dataset that includes \textit{Planck} low-$\ell$ CMB TTEE, high-$\ell$ CMB TTTEEE, CMB lensing, DESI DR1 BAO, Pantheon+ SNe Type Ia, local $H_0$ measurements, and DES Y1 weak lensing data. The Hubble tension is \REV{slightly reduced}\COMREV{relaxed} in 3-form field model\COMREV{ without any additional fine-tuning on the model parameters}, as the $H_0$ value derived from CMB + BAO dataset increase from $67.89\pm0.36{\rm km/s/Mpc}$ for $\Lambda$CDM model to $68.29^{+0.56}_{-0.61}{\rm km/s/Mpc}$ for the 3-form field model. As such, the model is \COMREV{statistically}\REV{modestly} favourable over $\Lambda$CDM when combining high-$z$ dataset of CMB and BAO with SNe and low-$z$ $H_0$ measurement. This does introduce tension between these datasets along $\Omega_{\rm m0}$ direction, a common price to pay among phantom dark energy model as the solution candidate to the Hubble tension. But as shown universally by statistical probes in this work, the lack of strong constraint on $\Omega_{\rm m0}$ makes it a reasonable trade-off. Overall, 3-form field model serves well as a theoretically motivated model for phantom dark energy.

However, the 3-form field does require a specific moment to exert its phantom-ness on the cosmological evolution, leading to a fine-tuning of the initial condition. This moment corresponds to $z \sim 1.7$, which has been pointed out in \cite{Akarsu:2021fol, Akarsu:2022typ, Akarsu:2023mfb, Bouhmadi-Lopez:2025ggl} as a potential time frame for the dark energy to suddenly become relevant. In addition, we show that the 3-form perturbation becomes stable at the same time, thus naturally \COMREV{suppressing the instability that may affect the CMB and matter power spectrum. This is confirmed by the $f\sigma_8$, CMB and matter power spectrum extracted from CAMB}\REV{setting it as an effective dark energy with a valid perturbation description once it dominates. However, we remind the reader that we have not incorporated the \HWC{correct }3-form perturbation into the Boltzmann code\HWC{~as it would inevitably lead to gradient instability}}
.

In conclusion, the 3-form field shows \COMREV{strong }promise in the cosmological setting. From the theory point of view, it has rich dynamics, including the existence of LSBR. From the observation point of view, the 3-form field is \COMREV{statistically}\REV{mildly} preferable over a cosmological constant\COMREV{, without inducing any instability on the perturbation} \REV{via rapid increase of dark energy density}. It therefore may serve as a candidate solution to the Hubble tension problem.

\appendix
\section{Centre manifold: method and explicit construction for our model}
\label{app:CM-ours}

\subsection{Method in a nutshell}

Standard linear stability analysis becomes inconclusive when a fixed point of a dynamical system has eigenvalues with zero real parts. In such cases, \emph{centre manifold theory} provides a systematic way to reduce the system to its essential degrees of freedom, restricted to the subspace associated with these neutral directions. The stability of the reduced dynamics then determines the stability of the full system~\cite{Carr:1981,Wiggins:1990}.

Consider the autonomous system
\begin{align}
  \dot{\mathbf{z}} = \mathbf{F}(\mathbf{z}),
  \label{cdy1}
\end{align}
where $\mathbf{F}$ is a smooth vector field on $\mathbb{R}^n$.  
Let $\mathbf{z}_0$ be a fixed point, $\mathbf{F}(\mathbf{z}_0)=0$.  
Introducing $\mathbf{v}=\mathbf{z}-\mathbf{z}_0$ gives the expansion
\begin{align}
  \dot{\mathbf{v}} = J\,\mathbf{v} + \mathbf{N}(\mathbf{v}),
  \label{cdy2}
\end{align}
with $J$ the Jacobian of $\mathbf{F}$ at $\mathbf{z}_0$ and $\mathbf{N}(\mathbf{v})$ collecting nonlinear terms.

The phase space $\mathbb{R}^n$ can be decomposed into the direct sum of three invariant subspaces:
\begin{align}
\mathbb{R}^n = \mathcal{S} \oplus \mathcal{U} \oplus \mathcal{C},
\end{align}
where $\mathcal{S}$, $\mathcal{U}$ and $\mathcal{C}$ correspond to the stable, unstable and centre subspaces, spanned by the eigenvectors of $J$ with negative, positive and zero real parts, respectively.  
If $\mathcal{U}$ is non-empty, the equilibrium is unstable by definition.  
When $\mathcal{U}$ is empty and at least one eigenvalue of $J$ has zero real part, the fixed point is \emph{non-hyperbolic}, and its stability must be analysed through centre manifold techniques.

After an appropriate linear transformation that separates the centre and stable directions, the dynamics near the equilibrium can be written as
\begin{subequations}
  \label{cmexvec}
  \begin{align}
    \dot{\mathbf{x}} &= A\,\mathbf{x} + \mathbf{f}(\mathbf{x},\mathbf{y}), \label{cenxdot}\\
    \dot{\mathbf{y}} &= B\,\mathbf{y} + \mathbf{g}(\mathbf{x},\mathbf{y}), \label{cenydot}
  \end{align}
\end{subequations}
where $(\mathbf{x},\mathbf{y}) \in \mathbb{R}^c \times \mathbb{R}^s$, with $c=\dim(\mathcal{C})$ and $s=\dim(\mathcal{S})$.  
Here, $A$ has eigenvalues with zero real parts, and $B$ has eigenvalues with strictly negative real parts.  
The functions $\mathbf{f}$ and $\mathbf{g}$ satisfy
\begin{align}
\mathbf{f}(0,0)=\mathbf{g}(0,0)=0, \qquad \nabla \mathbf{f}(0,0)=\nabla \mathbf{g}(0,0)=0.
\end{align}

\textit{Centre Manifold:} A geometrical space is a centre manifold for the system~\eqref{cmexvec} if it can be locally represented as
\begin{align}
  W^{c}(0) = \{\, (\mathbf{x},\mathbf{y}) \in \mathbb{R}^c \times \mathbb{R}^s \;|\;
  \mathbf{y} = \mathbf{h}(\mathbf{x}),\, |\mathbf{x}| < \delta,\, \mathbf{h}(0) = 0,\, \nabla \mathbf{h}(0) = 0 \,\},
\end{align}
for $\delta$ sufficiently small and $\mathbf{h}(\mathbf{x})$ a sufficiently regular mapping $\mathbb{R}^c \rightarrow \mathbb{R}^s$.

The conditions $\mathbf{h}(0)=0$ and $\nabla \mathbf{h}(0)=0$ ensure that the manifold $W^c(0)$ is tangent to the centre subspace $\mathcal{C}$ at the origin.  
The dynamics restricted to this manifold is governed by
\begin{align}
  \dot{\mathbf{u}} = A\,\mathbf{u} + \mathbf{f}\!\left(\mathbf{u}, \mathbf{h}(\mathbf{u})\right),
  \label{dynred}
\end{align}
for $\mathbf{u}\in\mathbb{R}^c$ sufficiently small.  
The stability properties of the reduced system~\eqref{dynred} determine the stability of the full system~\eqref{cmexvec}.

In general, the exact form of $\mathbf{h}(\mathbf{x})$ cannot be obtained analytically. However, this is not necessary for stability analysis.  
Differentiating $\mathbf{y}=\mathbf{h}(\mathbf{x})$ with respect to the independent variable gives $\dot{\mathbf{y}} = \nabla \mathbf{h}(\mathbf{x}) \cdot \dot{\mathbf{x}}$, and substituting Eqs.~\eqref{cenxdot}–\eqref{cenydot} leads to the \emph{invariance condition}
\begin{align}
  \nabla \mathbf{h}(\mathbf{x})\!\left[A\,\mathbf{x} + \mathbf{f}\!\left(\mathbf{x},\mathbf{h}(\mathbf{x})\right)\right]
  - B\,\mathbf{h}(\mathbf{x}) - \mathbf{g}\!\left(\mathbf{x},\mathbf{h}(\mathbf{x})\right) = 0,
  \label{neqn}
\end{align}
which defines $\mathbf{h}(\mathbf{x})$.  
Since only the local behaviour near the origin is reequired, one can expand $\mathbf{h}$ and determine the coefficients by matching powers in Eq.~\eqref{neqn}.  
This local approximation uniquely specifies the centre manifold to the desired order and suffices to infer the stability of non-hyperbolic equilibria.  
A more technical and rigorous exposition of all the statements presented here can be found in~\cite{Carr:1981}.

\subsection{Application to our model: linear split, centre manifold, and stability}
\label{app:ours-setup}

We focus on the positive branch of the non-hyperbolic equilibrium point,
\begin{align}
B^{+}:\qquad
(v,y,\mathfrak{h},r)=\left(\frac{\sqrt{5}-1}{2},\,1,\,1,\,0\right).
\end{align}
The analysis for the negative branch \(B^{-}\) follows identically and leads to the same stability conclusions.

To apply the centre manifold method, we first translate this fixed point to the origin by defining
\begin{align}
\tilde v = v - \frac{\sqrt{5}-1}{2}, \qquad
\tilde y = y - 1, \qquad
\tilde{\mathfrak{h}} = \mathfrak{h} - 1, \qquad
\tilde r = r,
\end{align}
so that \((\tilde v,\tilde y,\tilde{\mathfrak{h}},\tilde r)=(0,0,0,0)\) corresponds to the equilibrium.  
Linearising the system around the origin gives the Jacobian
\begin{align}
J =
\begin{pmatrix}
-3 & -\tfrac{3}{10}(-5+\sqrt{5}) & 0 & 0\\[3pt]
0 & -3 & \tfrac{3}{2}e^{-\xi/9}\bar{V} & 0\\[3pt]
0 & 0 & 0 & 0\\[3pt]
0 & 0 & 0 & -2
\end{pmatrix},
\end{align}
whose eigenvalues and eigenvectors are, respectively,
\begin{align}
\{\lambda_i\} &= \{\,0,\,-3,\,-3,\,-2\,\},\\[2pt]
\{\mathbf{e}_i\} &= 
\Bigl\{
\bigl(-\tfrac{1}{20}(-5+\sqrt{5})e^{-\xi/9}\bar{V},\,\tfrac{1}{2}e^{-\xi/9}\bar{V},\,1,\,0\bigr),\;
(1,0,0,0),\;
(0,0,0,0),\;
(0,0,0,1)
\Bigr\}.
\end{align}
The single zero eigenvalue confirms the existence of a \emph{one-dimensional centre subspace}, spanned by
\begin{align}
\mathbf{e}_c \propto 
\bigl(-\tfrac{1}{20}(-5+\sqrt{5})e^{-\xi/9}\bar{V},\,\tfrac{1}{2}e^{-\xi/9}\bar{V},\,1,\,0\bigr),
\end{align}
while the remaining three eigenvectors correspond to stable modes with negative real parts.

Since the Jacobian has a vanishing eigenvalue, it is \emph{not diagonalisable}.  
Nevertheless, it can be expressed in the block form required for the application of centre manifold theory, separating the centre and stable dynamics.  
This is achieved by a further linear transformation that mixes the variables \((\tilde v,\tilde y,\tilde{\mathfrak{h}},\tilde r)\) according to the eigenstructure above.  
The transformation reads
\begin{align}
v_{\mathrm{CM}} &= \tilde v + \tfrac{1}{20}(-5+\sqrt{5})e^{-\xi/9}\bar{V}\,\tilde{\mathfrak{h}}, \nonumber\\
y_{\mathrm{CM}} &= -\tfrac{3(-5+\sqrt{5})}{10}\!\left[\tilde y - \tfrac{1}{2}e^{-\xi/9}\bar{V}\,\tilde{\mathfrak{h}}\right], \nonumber\\
\mathfrak{h}_{\mathrm{CM}} &= \tilde r, \nonumber\\ 
r_{\mathrm{CM}} &= \tilde{\mathfrak{h}}.
\end{align}
This linear map aligns the system with its natural invariant subspaces and isolates the centre variable.  
After this transformation, the system assumes the canonical structure reequired by the centre manifold theorem,
\begin{subequations}
\label{eq:CM-canonical}
\begin{align}
  \dot{u} &= f(u,\mathbf{y}),\\[2pt]
  \dot{\mathbf{y}} &= B\,\mathbf{y} + \mathbf{g}(u,\mathbf{y}),
\end{align}
\end{subequations}
where \(u=r_{\mathrm{CM}}\) parameterises the one–dimensional centre subspace (associated with the zero eigenvalue), and \(\mathbf{y}=(v_{\mathrm{CM}},y_{\mathrm{CM}},\mathfrak{h}_{\mathrm{CM}})\in\mathbb{R}^3\) collects the stable modes. In these coordinates, the stable block \(B\) takes the simple upper-triangular form
\begin{align}
B =
\begin{pmatrix}
-3 & 1 & 0 \\[3pt]
0 & -3 & 0 \\[3pt]
0 & 0 & -2
\end{pmatrix},
\end{align}
whose eigenvalues are \((-3,-3,-2)\).  
The off–diagonal entry couples the first two stable variables and prevents diagonalisation. The explicit forms of \(f(u,\mathbf{y})\) and \(\mathbf{g}(u,\mathbf{y})\) contain numerous high–order mixed nonlinearities and are therefore omitted here for brevity.  
All algebraic manipulations and the numerical evaluation of the reequired coefficients have been carried out using \textsc{Mathematica}~\cite{Mathematica}.

Having brought the system to the canonical form~\eqref{eq:CM-canonical}, we now construct the local centre manifold explicitly.  
The stable variables are expressed as smooth functions of the centre variable,
\begin{align}
  \mathbf{y}=\mathbf{h}(u), \qquad \mathbf{h}(0)=0,\quad \nabla \mathbf{h}(0)=0.
\end{align}
Expanding \(\mathbf{h}(u)\) near the origin and retaining terms up to third order gives
\begin{align}
  \mathbf{h}(u)=
  \begin{pmatrix}
    a_1 u^2 + a_2 u^3 + \mathcal{O}(u^4)\\[3pt]
    a_3 u^2 + a_4 u^3 + \mathcal{O}(u^4)\\[3pt]
    a_5 u^2 + a_6 u^3 + \mathcal{O}(u^4)
  \end{pmatrix},
  \label{eq:h-series-ours}
\end{align}
where the coefficients \(a_i=a_i(\bar{V},\xi)\) are obtained by substituting~\eqref{eq:h-series-ours} into the invariance condition~\eqref{neqn} and matching powers of~\(u\).  
Solving order by order yields
\begin{align}
  a_1 &= \frac{e^{-2\xi/9} \bar{V}}{600\,(-7+3\sqrt{5})}
  \Big[\,60(25-11\sqrt{5})\,e^{\xi/9}
        + \bar{V}\big(1215-537\sqrt{5} + (500-220\sqrt{5})\,\xi\big)\Big],\\[4pt]
  a_2 &= \frac{e^{-\xi/3} \bar{V}}{32400\,(-1165+521\sqrt{5})}
  \Big[\,16200(-843+377\sqrt{5})\,e^{2\xi/9}
        + \bar{V}\big(
              81(-119963+53649\sqrt{5})\,\bar{V} \nonumber\\[-2pt]
      &\hspace{3.2cm}
            + 540\,e^{\xi/9}\big(-41061+18363\sqrt{5}
               + 20(-843+377\sqrt{5})\,\xi\big) \nonumber\\[-2pt]
      &\hspace{3.2cm}
            + 20\,\bar{V}\,\xi\big(-217809+97407\sqrt{5}
               + 175(-843+377\sqrt{5})\,\xi\big)
           \big)\Big],\\[4pt]
  a_3 &= \frac{(-5+\sqrt{5})}{240}\,e^{-2\xi/9} \bar{V}
        \Big[\,36 e^{\xi/9} + \bar{V}\,(9+4\xi)\Big],\\[4pt]
  a_4 &= -\,\frac{(-5+\sqrt{5})}{4320}\,e^{-\xi/3} \bar{V}
        \Big[\,648 e^{2\xi/9}
             + \bar{V}\big(36 e^{\xi/9}(9+4\xi)
                       + \bar{V}\,(81 + 4\xi(9+7\xi))\big)\Big],\\[4pt]
  a_5 &= a_6 = 0.
\end{align}
Higher–order terms (\(u^4\) and beyond) are unnecessary for determining stability.

Substituting \(\mathbf{h}(u)\) into the reduced equation for the centre variable yields
\begin{align}
  \dot u \;=\; \frac{2}{3}\,e^{-\xi/9}\,\bar{V}\,\xi\,u^2
  \;+\; \mathcal{O}(u^3),
  \label{eq:CM-reduced-ours}
\end{align}
which fully captures the dynamics near the equilibrium.  
Equation~\eqref{eq:CM-reduced-ours} provides the normal form of the dynamics on the centre manifold.  
The quadratic leading term shows that the fixed point is \emph{semistable}: for a generic system, trajectories with \(u>0\) move away from the origin (unstable to the right), while those with \(u<0\) approach it (stable from the left) when \(\bar{V}\xi>0\); the opposite holds when \(\bar{V}\xi<0\).  

In our case, however, the centre variable corresponds to \(u=r_{\mathrm{CM}}=\tilde{\mathfrak{h}}=\mathfrak{h}-1\), where \(0\le\mathfrak{h}\le1\), and hence \(u\le0\).  
Moreover, since we assume \(\bar{V}>0\) and \(\xi>0\), the coefficient of the quadratic term in~\eqref{eq:CM-reduced-ours} is strictly positive.  
Consequently, for all physically admissible states \(u<0\), we have \(\dot{u}>0\), implying that \(u(t)\) increases monotonically toward the equilibrium \(u=0\).  
Therefore, within the physical domain, the fixed point \(B^{+}\) acts as an \emph{attractor along the centre manifold}.  
The negative branch \(B^{-}\) exhibits the same qualitative behaviour under the same physical restriction.

\section{Statistical probes}
\label{sec:stat}

The definition of each probe is listed below:
\begin{align}
\ln B (D|M) &\equiv \ln \int P(D|\theta_M) P(\theta_M) d\theta_M = -\ln V_M - \ln \left< \left( P(D|\theta_M) \right)^{-1} \right>_{M|D}  \label{eq:BE}  \,,\\
{\rm DIC} (D|M) &\equiv \ln P(D|\theta_{M,\,map}) - 2F(D|M)  \,,\\
{\rm WAIC} (D|M) &\equiv - F(D|M) + {\rm BMD} (D|M) / 2  \,,\\
F(D|M) &\equiv \left< \ln P(D|\theta_M) \right>_{M|D} \equiv \ln B (D|M) + {\rm KL} (D|M)  \,,\\
{\rm BMD} (D|M) &\equiv 2 \left( \left< \left( \ln P(D|\theta_M) \right)^2 \right>_{M|D} - \left< \ln P(D|\theta_M) \right>_{M|D}^2 \right)  \,,\\
-\ln R (D_1, D_2 | M) &\equiv - \ln B (D_1 D_2 | M) + \ln B (D_1 | M) + \ln B (D_2 | M)  \,,\\
{\rm GoF} (D_1, D_2 | M) &\equiv - \ln P(D_1 D_2 |\theta_{M,\,map}) + \ln P(D_1 |\theta_{M,\,map}) + \ln P(D_2 |\theta_{M,\,map})  \,,\\
S (D_1, D_2 | M) &\equiv - F(D_1 D_2 | M) + F (D_1 | M) + F (D_2 | M)  \,,
\end{align}
where $D$ and $M$ denote the dataset and the model, $\theta_M$ are the model parameters, $P(D|\theta_M)$, $P(\theta_M)$, $P(\theta_M|D) \equiv P(D|\theta_M) P(\theta_M) / B(D|M)$ are the likelihood, prior probability and posterior probability, $\left< ... \right>_{M|D} = \int ... P(\theta_M|D) d\theta_M$ is the MCMC mean, $V_M \equiv \int P(\theta_M) d\theta_M$ is the prior volume, $F$ is the deviance, ${\rm KL} (D|M) = \left< \ln P(\theta_M|D) - \ln P(\theta_M) \right>_{M|D}$ is the Kullback-Leibler divergence, BMD is the abbreviation of Bayesian model dimension, and $map$ is the abbreviation of maximum-a-posteriori, i.e., the Bayesian estimate of the model parameters.

While probes such as DIC, WAIC and Bayesian ratio can be easily interpreted using Jeffreys’ scale as shown in table~\ref{table:jeffreys}
, the goodness of fit and suspiciousness require further processing. As these two probes follow the BMD-dimensional $\chi^2$ distribution \cite{PhysRevD.100.023512}, we may convert a value $p$ into traditional $\sigma$ value as ${\rm CDF}_1^{-1} \left( {\rm CDF}_{\rm BMD} \left( \sqrt{{\rm BMD} + 2p} \right) \right)$ where ${\rm CDF}_d (p) = \int_0^p e^{-\chi^2}\chi^{2d - 2}d\chi^2$ is the cumulative distribution function of a $d$-dimensional Gaussian distribution.

\begin{table}[]
\centering
\begin{tabular}{|c|c|}
\hline
$\ln B (M_2) - \ln B (M_1)$, $\ln R$, etc.    & Interpretation    \\
\hline
$>5$        & Strongly disfavored / tensioned   \\
$2.5\sim5$  & Moderately disfavored / tensioned \\
$1\sim2.5$  & Weakly disfavored / tensioned     \\
$-1\sim1$   & Inconlusive                       \\
$-2.5\sim-1$& Weakly favored / aligned          \\
$-5\sim-2.5$& Moderately favored / aligned      \\
$<-5$       & Strongly favored / aligned        \\
\hline
\end{tabular}
\caption{\label{table:jeffreys}
Jeffreys' scale for deciding the evidence of model $M_1$ over $M_2$ or the tension between datasets.}
\end{table}

\acknowledgments
M. B.-L. is supported by the Basque Foundation of Science Ikerbasque. Our work is supported by the Spanish Grant PID2023-149016NB-I00 (MINECO/AEI/FEDER, UE). This work is also supported by the Basque government Grant No. IT1628-22 (Spain). 
HWC is supported by the NSFC Grant No.~12250410250 and No.~12347133. 
C. G. B. acknowledges financial support from the FPI fellowship PRE2021-100340 of the Spanish Ministry of Science, Innovation and Universities. M.B.-L. is grateful to the hospitality of LeCosPA, NTU (Taiwan) where this work was initiated during one of her visits.  This article
is based upon work from the COST Action CA21136
“Addressing observational tensions in cosmology with
systematics and fundamental physics” (CosmoVerse),
supported by COST (European Cooperation in Science
and Technology).

\bibliographystyle{JHEP}

\bibliography{bibliography.bib}

\end{document}